\documentclass[amsmath,amssymb,prl,superscriptaddress,nofootinbib]{revtex4}
\usepackage{graphicx}
\usepackage{dcolumn}
\usepackage{bm}
\usepackage{booktabs}

\def\d{{\rm d}}

\begin{document}

\markboth{Peter K. S. Dunsby and Orlando Luongo}{\,}
\title{On the theory and applications of modern cosmography}

\author{Peter K. S. Dunsby}
\affiliation{Department of Mathematics and Applied Mathematics, University of Cape Town, South Africa, Rondebosch 7701, Cape Town, South Africa.}
\affiliation{Astrophysics, Cosmology and Gravity Centre (ACGC), University of Cape Town, Rondebosch 7701, Cape Town, South Africa.}
\affiliation{South African Astronomical Observatory,  Observatory 7925, Cape Town, South Africa.}

\author{Orlando Luongo}
\affiliation{Department of Mathematics and Applied Mathematics, University of Cape Town, South Africa, Rondebosch 7701, Cape Town, South Africa.}
\affiliation{Astrophysics, Cosmology and Gravity Centre (ACGC), University of Cape Town, Rondebosch 7701, Cape Town, South Africa.}
\affiliation{Istituto Nazionale di Fisica Nucleare (INFN), Sezione di Napoli, Via Cinthia, I-80126 Napoli, Italy.}

\begin{abstract}
Cosmography represents an important branch of cosmology which aims to describe the universe without the need of postulating  \emph{a priori} any particular cosmological model. All quantities of interest are expanded as a Taylor series around here and now, providing in principle, a way of directly matching with cosmological data. In this way, cosmography can be regarded a model-independent technique, able to fix cosmic bounds, although several issues limit its use in various model reconstructions. The main purpose of this review is to focus on the key features of cosmography, emphasising both the strategy for obtaining the observable cosmographic series and pointing out any drawbacks which might plague the standard cosmographic treatment. In doing so, we relate cosmography to the most relevant cosmological quantities and to several dark energy models. We also investigate whether cosmography is able to provide information about the form of the cosmological expansion history, discussing how to reproduce the dark fluid from the cosmographic sound speed. Following this, we discuss limits on cosmographic priors and focus on how to experimentally treat cosmographic expansions. Finally, we present some of the latest developments of the cosmographic method, reviewing the use of rational approximations, based on cosmographic Pad\'e polynomials. Future prospects leading to more accurate cosmographic results, able to better reproduce the expansion history of the universe are also discussed in detail.
\end{abstract}

\maketitle

\tableofcontents

\newpage

\section{Introduction}\label{introduzione}

Assuming that the geometry of the universe on large scales is well described by the Robertson-Walker (RW) metric, a wide number of robust observations have now placed restrictions on the cosmological dynamics, leading to the conclusion that the universe is inescapably speeding up \cite{riess,perlmutter}. This phenomenon cannot be explained if the cosmic matter budget is made up only from contributions from baryonic and cold dark matter, since their contribution to the RHS of Einstein's equations lead to a decelerated expansion rate, rather than the observed acceleration \cite{miao}. From a theoretical point of view, a further component, different from standard matter, is thought to be responsible for present-time dynamics. Its contribution enters Einstein's equations and drives the cosmic evolution \cite{copeland}. The nature of this additional component is is the subject of a significant number of investigations and induces a kind of \emph{weak inflation} which persists at present time, accelerating the expansion rate of the universe today. Several explanations of this phenomena are possible and for a RW metric, two solutions to this problem have received extensive study. From the one side, the first class of models involves extensions of general relativity, via more complicated descriptions of the gravitational action \cite{extended}. From the other side, some type of exotic (perfect) fluid may represent a \emph{source} for the above-quoted mysterious dark component. Any conceivable source able to accelerate the universe needs a negative equation of state, able to counterbalance the gravitational attraction\footnote{For a different perspective, in which gravity may exhibit regions of gravitational repulsion see \cite{ioequevedo}.}. In a homogeneous and isotropic universe the fluid responsible for the dynamics of the universe is dubbed \emph{dark energy} and its evolution is the subject of much current research \cite{darkenergy}.

The name \emph{dark} relates to our inability to describe its nature, i.e., a measure of our ignorance in defining its fundamental physical properties. If the cosmic density driving the acceleration is assumed to be constant, one recovers the case of the \emph{cosmological constant} $\Lambda$. Its physical interpretation is essentially related to the zero-point energy of quantum field theory. The corresponding description of the universe, known as the $\Lambda$CDM model, has reached the status of the \emph{concordance model}. However, although suggestive and simple, it is severely plagued by a number of conceptual issues which have led to doubts that it is the definitive paradigm \cite{ilmodello}. Another perspective, in analogy with the $\Lambda$CDM model, is to assume a constant pressure and an evolving dark energy density. This approach is highly adaptable to observations and shows the important property that dark energy and matter become emergent fluids \cite{ilmiopaperconquew}. The approach is sometimes referred to as the \emph{dark fluid} and degenerates  with the standard cosmological model by having a variable barotropic factor \cite{darkfluid}. The two approaches are almost indistinguishable even at a fundamental level and only direct measurements of the equation of state would discriminate whether the $\Lambda$CDM model is favoured over extended gravity or a dark fluid. For these reasons, the need of model independent techniques is essential in order to describe the expansion of the universe, without postulating the dark term \emph{a priori}. Alternatively speaking, combined model independent measures would break the degeneracy among cosmological coefficients. This strategy would indicate whether dark energy evolution took place in the past or whether the dark energy density corresponds to a pure vacuum energy cosmological constant \cite{losdos}.

Among all reasonable approaches, the so-called cosmography of the universe has recently attracted a lot of attention \cite{iniziocosmografia,harrison}. Here we review the principal aspects of cosmography and show how one applies cosmography to the problem of determining the correct dark energy evolution. To do so, we summarise the principal advantages and the main techniques of cosmographic reconstructions. We define the set of coefficients that determine the cosmographic series, and we relate cosmography to other cosmological observables. Furthermore, we describe several consequences of cosmography in the areas of thermodynamics and dark energy evolution respectively. We therefore show how the cosmographic series changes if the minimal cosmological model is modified and demonstrate how to obtain physical priors on the cosmographic coefficients. We also highlight several disadvantages of considering cosmography as a pure model-independent technique if some precise conditions do not hold. Thus, we give high consideration to the caveat of present-time data, which dampens the enthusiasm towards the use of cosmographic. Finally, we underline the modern use of refined approximations which can be performed in the context of cosmography by means of rational polynomial functions. We also demonstrate what one expect for \emph{future-cosmographic approachs}, capable of framing the universe's expansion history through direct measurements of the cosmographic series beyond present limits.

The review is structured as follows: in the next section, we describe the connection between the Friedmann-Robertson-Walker universe and the cosmographic coefficients. To do so, we show what the easiest way to handle the cosmographic expansion is and we demonstrate how to relate cosmological quantities to the cosmographic series. In Sec. III, we rewrite the Hubble rate, the pressure, the density and the luminosity distances in terms of cosmographic coefficients. To the best of our knowledge, we also propose a way to reproduce the most popular cosmological distances in terms of the cosmographic series. Furthermore, we review the dark fluid approach and the way to obtain it from cosmography, by considering a cosmographic sound speed. We also demonstrate how to define a cosmographic equation of state. In Sec. IV, we describe a scheme to get the cosmographic series for several cosmological models of particular interest, focusing on the $\Lambda$CDM and $\omega$CDM models and on the Chevallier-Polarski-Linder parametrisation. We also propose how to treat the cosmographic coefficients for a generic model with an additive dark energy term. In Sec. V, we highlight the problems of cosmography, showing that present-data is not enough to guarantee precise cosmographic bounds and how more generally the cosmographic treatment can be misleading. We provide an explanation for these issues and we attempt to give some hints to alleviate these problems. In Sec. VI, we go beyond the standard cosmographic treatment and we emphasise the role played by rational approximations in the cosmographic regime. We describe the modern Pad\'e approximation of cosmographic observables and we briefly describe the perspectives toward a high-redshift cosmography which goes beyond the standard recipe. Finally, in Sec. VII, we summarise the basic points of our review and we propose future developments and perspectives of the cosmographic approach.

\section{The path to cosmography: basic requirements and strategies}

Cosmography represents a general strategy for dealing with the cosmological parameters in terms of the kinematics of the universe only. Thus, the cosmographic treatment has the nice property of being model-independent and therefore does not depend on \emph{any} particular dark energy model. Consequently, the energy momentum tensor is not defined \emph{a priori}, i.e., one does not need to postulate the equation of state to determine the dynamics of the universe \cite{noein}. If one is able to quantify the cosmographic constraints at higher redshifts and extend cosmography to the early universe, it would be possible to reconstruct the dark energy contribution, instead of postulating its form at the very beginning in Einstein's equations \cite{noein2}. However, the present state of the art basically allows one to focus on the late stages of the universe's evolution. Typically, all Taylor expansions associated with cosmography lie in the observable domain $z\ll1$ and enable one to fix limits on the current universe. In what follows, we highlight the key aspects of \emph{late-time cosmography} \cite{noein3}. To do so, we discuss the basic requirements of cosmography and hence suggest ways of how one might go about formulating the theory of \emph{high redshift cosmography}. So, in lieu of considering cosmography as a simple treatment to fix bounds on observable quantities, we present a self-consistent \emph{cosmographic method} and discuss how to relate it to reconstructing dark energy models.

The first step is to define suitable cosmographic coefficients. In order to achieve this, we first assume that the energy budget of the universe is built up from a set of perfect fluids \cite{fluidi}. In a homogeneous and isotropic universe\footnote{Cosmography beyond this hypothesis is still object of debate. See for example \cite{vi2005}.}, this leads to the Friedmann equations\footnote{Hereafter in units $8\pi G=c=1$.}:
\begin{equation}\label{frweqs}
H^2=\frac{1}{3}\rho-\frac{k}{a^2}\,,\qquad\quad
\dot H+H^2 = -\frac{1}{6}\left(3P+\rho\right)\,,
\end{equation}
which describe the expansion of the universe, once an equation of state $P=P(\rho)$ is given. In particular, the first Friedmann equation tells us that all cosmological densities contained in $\rho$, are connected to $H^2$. This property can be rephrased, by noticing that the first Friedmann equation represents a \emph{constraint}, since it does not involve any time derivatives of $H(t)$. In contrast, the second Friedmann equation contains the term $\dot H$, which describes the sign of the acceleration of the universe and correspondingly its dynamics. Combining the above expressions, we infer the continuity equation $\dot{\rho}+3H(\rho+P)=0$. This expression can be alternatively obtained from the twice contracted Bianchi Identities $\nabla_\nu T^{\mu\nu}=0$. Moreover, it provides the formal solution $\rho(z)=C\,e^{\int_0^z\frac{1+\omega(\xi)}{1+\xi}\d\xi}$, expressed in terms of the redshift $z$, where we have made use of the identity $\frac{dz}{dt}=-(1+z)H(z)$, with $C$ an integration constant. Thus, we have:
\begin{equation}\label{conti2}
\frac{d\rho}{dz}=3\frac{P+\rho}{1+z}\,,
\end{equation}
which represents an alternative and sometimes more practical form of the continuity equation as a function of the redshift $z$. In our picture, $\rho(z)$ stands for the total density of a generic combination of perfect fluids. Due to the linearity of Eq. (\ref{conti2}), this quantity can be split into a sum of fluid densities, representing the constituents of the whole energy momentum tensor, i.e., $\rho=\sum_{i=0}^{\infty}\rho_i(z)$. The densities $\rho_i$ and $\rho$ are defined once the corresponding equations of state  $\omega_i(z)$ and $\omega(z)$ for the fluid are known. In general, one assumes the Friedmann equations to hold throughout the entire evolution of the universe and so, in order to find out the dark energy contribution, it is necessary to determine the barotropic equation of state for the pressure. In other words, one needs the functional dependence of $P$ on the density $\rho$ or more practically as a function of the redshift $z$\footnote{Notice that the total equation of state of the universe $\omega$ does not differ (in form) from the equations of state of single components, i.e., $P_i=\omega_i\rho_i$. However, in such cases $\omega_i\neq\omega$, since for each species $i$, $\omega_i\equiv{P_i\over\rho_i}$, whereas $\omega\equiv{\sum_iP\over\sum_i\rho_i}$, since $P\equiv\sum_iP_i$ and $\rho\equiv\sum_i\rho_i$. The hypothesis of barotropic fluids assumes therefore that the pressure, and correspondingly the equation of state, does not depend upon the entropy $S$.}. From this hypothesis, one is able to obtain the expansion of cosmological quantities of interest. In particular, by considering the total pressure, representing a source of the acceleration of the universe, one has:
\begin{equation}
P \approx P_0 + \kappa_0^{(1)} (\rho-\rho_0) + \frac{1}{2}\kappa^{(2)}_{0}(\rho-\rho_0)^2 + \ldots\;,
\end{equation}
where $\kappa^{(n)}_{0}\equiv\frac{d^n P }{d\rho^n}\Big|_{0}$ is evaluated at $z=0$ and is truncated\footnote{Employing barotropic fluids and the above Taylor expansion represents a technique which is certainly valid in cases of adiabatic expansion. When entropy fluctuations become important, it needs to be modified.} at $n=2$.

From now on, we only employ barotropic fluids with $\Omega\equiv\frac{\rho}{\rho_c}$ defined to be the (normalised) total density of the universe, setting the critical density $\rho_c\equiv3H_0^2$ in our units. Late time observations appear to constrain $\Omega\simeq1$. More stringent bounds may be also found at different stages of universe's evolution, i.e., $|\Omega-1| < {\cal O}(10^{-16})$ during Nucleosynthesis or $|\Omega-1| < {\cal O}(10^{-64})$ in the Planck era \cite{ade}. Hence, the universe may be approximated by a vanishing spatial curvature and negligible contributions coming from neutrinos and radiation today. From now on we simplify our cosmographic treatment by characterising the matter content of the universe as a mixture of pressure-less matter (baryons and cold dark matter) and dark energy only, neglecting any additional contributions (neutrinos, spatial curvature and radiation). Later on, we consider the role of spatial curvature, dealing with its consequences on the cosmographic dark energy evolution. The ability to  neglect all densities except matter and dark energy comes from the fact that we focus only on the late time universe. In such a picture, cosmography works well in the regime where $H_0 a_0/c \gg 1$, which represents the condition for obtaining an accelerated expansion today

\footnote{Although we expand the cosmographic coefficients in our epoch,
this condition also represents a basic requirement for obtaining the onset
of inflation. In other words, one can think that the cosmographic series (if extended all the way back to the early universe) is even valid during inflation. This fact, however, does not suggest that $k=0$ during inflation, but seems to suggest that it should be small, as the standard model effectively provides.}, with $a(t_0)=a_0$. In this picture, the expansion of the pressure in terms of $\rho$ is a first example of a cosmographic series. However, it most likely represents a unsuitable way of proceeding, since it is generally difficult to compare the pressure \emph{directly} with cosmic data. Moreover, it is not clear to expand the pressure, or in general other observable quantities, so the easiest choice is to obtain a Taylor expansion of the scale factor $a(t)$ around our time, $t=t_0$\footnote{In the next section, we will provide more details on how to place significant constraints on the observable quantities, by expanding $a(t)$.}.

Assuming the cosmic time $t$ as suitable variable, the expansion of $a(t)$ reads:
\begin{equation}\label{serie1a}
a(t)=\lim_{\mathcal N\rightarrow\infty}\sum_{h=0}^{\mathcal N}{a^{(h)}\over h!}\bar t^h\,,
\end{equation}
where $a^{(h)}\equiv\frac{\d^h a}{\d t^h}$ and $\bar t\equiv t-t_0>0$ \footnote{The interval $\bar t$ should be positive to guarantee causality between emitted and observed light. In other words, the series takes in to account that a traveling photon starts from a point and is detected at $t$, with $t-t_0>0$.}.

In particular, without applying the limit definition, one truncates at arbitrary orders, leading to approximations which hold up to $\mathcal N$ and plausibly remain relevant as long as the $\mathcal N+1$ term is less than $\mathcal N$, normalised by (at least) a factor 10, having:
\begin{equation}\label{factor}
\bar t \leq \frac{a^{(\mathcal N)}(0)(\mathcal N+1)}{10a^{(\mathcal N+1)}(0)}\,.
\end{equation}
The scale factor is essentially a non-observable quantity, as discussed above, however it provides a nice outcome, since the cosmographic series naturally arises once causality is guaranteed. In this scheme, a commonly used convention is to define the following set of cosmographic coefficients in the  \emph{cosmographic series}:
\begin{subequations}\label{CSdef}
\begin{align}
H \equiv \frac{1}{a} \frac{\d a}{\d t}\,,\quad q \equiv -\frac{1}{a H^2} \frac{\d^2a}{\d t^2}\,,\quad j  \equiv \frac{1}{a H^3} \frac{\d^3a}{\d t^3}\,,\\
s \equiv \frac{1}{a H^4} \frac{\d^4a}{\d t^4}\,,\quad l  \equiv \frac{1}{a H^5} \frac{\d^5a}{\d t^5}\,,\quad m \equiv \frac{1}{a H^6} \frac{\d^6a}{\d t^6}\,.
\end{align}
\end{subequations}
These terms can be constrained by observations and can be related to any observable quantities in cosmology. The coefficients represent a \emph{cosmographic model constructed from a set of independent quantities} and
do not depend upon the form of the dark energy fluid for the simple reason that they are not functions of either $\omega$ or $\omega_i$. The terms entering the cosmographic series are referred to as the \emph{Hubble parameter} ($H$), the \emph{acceleration and jerk}\footnote{In the literature, it is still common to use \emph{deceleration} instead of acceleration, because before the discovery of dark energy $q$ was thought to be positive, thus indicating a deceleration. Moreover, its variation associated with $j$ was previously named \emph{surge} by \cite{harrison}, in which the physical meaning of $j$ was quite different from the modern interpretation.} ($q$ and $j$), the \emph{snap and lerk} ($s$ and $l$) and the \emph{max-out} parameter $m$. In theory, these coefficients, when evaluated at the present time, become effective observables from which one can construct the low-$z$ expansion history of the universe. More precisely, we can  write the scale factor in terms of the present time cosmographic series:
\begin{equation}
a(t)   \sim  1+  H_0 \Delta t - \frac{q_0}{2}  H_0^2\Delta t^2+\frac{j_0}{6} H_0^3 \Delta t^3 +   \frac{s_0}{24}  H_0^4\Delta t^4 +\frac{l_0}{120}  H_0^5\Delta t^5\;,
\end{equation}
encoding the expansion history directly through the measurable cosmographic coefficients. From the shape of the Hubble curve, it is possible to deduce the physics behind each coefficient. In particular, throughout the entire evolution of the universe, the sign of $q$ indicates whether the dynamics is accelerated or decelerated. In other words, a positive acceleration parameter indicates that standard gravity predominates over the other species, whereas a negative sign provides a repulsive effect which overcomes the standard attraction due to gravity.

All of this aside, the acceleration parameter alone does not account for the whole dynamics. To determine if it changes sign during its evolution, the sign of $j$ is a key observable. For example, a positive jerk parameter would indicate that there exists a transition time when the universe modifies its expansion. Corresponding to this transition, the $q$ modulus tends to zero and then changes its sign. The two terms, i.e., $q$ and $j$ fix the local dynamics, but are not enough to discriminate between a cosmological model that admits an evolving dark energy term or one with a pure cosmological constant \cite{noosi}. In order to determine if there is any evolution of dark energy, the value of $s$ is strictly necessary.  Departures from the expected value of $s$, evaluated in the concordance model, affects the functional dependence of dark energy on the redshift $z$, indicating that it evolves as the universe expands. In particular, at higher redshifts, the terms $s$ and $l$ both influence higher orders of the Taylor expansion. From these considerations one can handle two \emph{cosmographic regimes}: the lower one valid up to a redshift $z\leq2$ and the higher valid for $z\geq2$. For example, if $z\leq1$, one does not need the complete list of cosmographic coefficients in order to determine the shape of the Hubble curve. Indeed, to  completely fix the cosmographic series, it is enough to put bounds on $H,q,j$, neglecting the other terms. This coarse-grained approach fails for $z\geq 1$ up to $z\simeq 2$, where one even needs to fix $s$, $l$. Additional higher orders also require $m$. Hence, depending on the particular data sets involved in the statistical analysis, one can focus attention on the broadening of cosmographic coefficients. The variation of $s$, for example, is essentially due to the sign of the lerk parameter and indicates how the shape becomes smooth at redshifts $z\geq1$. Although a precise physical interpretation of such coefficients has been built up above, present data is not enough to guarantee overwhelming constraints on those terms \cite{nohq}. High precision treatments are essentially limited by the particular set of data involved in the analysis. Typically, if one considers one single set of cosmic data, more often than not, one obtains systematics which do not allow one to put accurate limits on the cosmographic series. As it will be clarified later, a combined analyses is required to obtain a cosmographic series with lower systematic errors. For example, employing the union 2.1 compilation of presently available supernova measurements, confines data in the range $z\leq1.414$. So, with this data it appears pointless to constrain $s,l,m$ or to expect to find accurate bounds on them, because their meaning would only be clearer at $z\geq2$. On the other hand, by invoking the differential age independent $H$ measurements, one may argue that those coefficients are more significative, since $z>2$ \cite{ohd}. Thus, the choice of one data set over  another becomes essential in order to obtain the correct cosmographic series at arbitrary order. In general, a viable recipe would be to combine cosmic data to get tighter bounds, which make use of low and high redshift regimes.

The last considerations even jeopardise the possibility of obtaining theoretical priors on the cosmographic series. In fact, to manifestly define the correct intervals for which the cosmographic coefficients span, a significative advantage would consist of intertwining among them all the coefficients. To this end, we have:
\begin{subequations}
\label{jh35kdj}
\begin{align}
\frac{\d q}{\d t}&=-H(j-2q^2-q)\,,\\
\frac{\d j}{\d t}&=H\big[s+j(2+3q)\big]\,,\\
\frac{\d s}{\d t}&=H\big[l+s(3+4q)\big]\,,\\
\frac{\d l}{\d t}&=H\big[m + l(4+5q)\big]\,.
\end{align}
\end{subequations}
These expressions allow one to fix the corresponding limits over the $N$-coefficient, while the $N-1$ terms are known. Moreover, coming back to the case of the total pressure, we get
\begin{subequations} \label{eq:pressureandD}
\begin{align}
P& = -H^2 \left(1-2q\right)\,,\\
\frac{dP}{dt} & = -2H^3 \left(j-1\right)\,,\\
\frac{d^2P}{dt^2} &= -2H^4 \Big[s-j+3q+3\Big]\,, \\
\frac{d^3P}{dt^3}  &= -2H^5 \Big[ (l-j(1+q)-3q(7+2q)-2(6+s))\Big]\,,\\
 \frac{d^4P}{dt^4}  &= -2H^6\left\{m - 3l + j^2 + 12j(2+q)+3\Big[20+s+q(48 + q(27+2q)+s\Big]\right\}\,,
\end{align}
\end{subequations}
and using the Friedmann equations, we have:
\begin{subequations} \label{densitytutte}
\begin{align}
\frac{d\rho}{dt} & =-6H^3(q+1)\,,\\
\frac{d^2\rho}{dt^2}&=6H^4\Big[j+q(q+5)+3\Big]\,,\\
\frac{d^3\rho}{dt^3}&=6H^5\Big[s-j(7+3q)-3q(9+4q)-12\Big]\,,\\
\frac{d^4\rho}{dt^4}&=6H^6\left\{l+3j^2+ j(48+44q)+q\big[3q(39+4q)-4(s-42)\big]-9s+60\right\}\,,\\
\frac{d^5\rho}{dt^5}&=6H^7\Big\{m-l(11+5q)-50j^2-30\big[12+q(2+q)(20+9q)\big]-10j\big[36+q(51+8q)-s\big]-5(15+14q)s\Big\}\,.
\end{align}
\end{subequations}
The corresponding transition redshift $z_{tr}$ represents the redshift which corresponds to the transition between a matter dominated to a dark energy dominated universe. Summing up, in the redshift domain $z\leq1$, $j_0$ shows a change of slope of the whole dynamics of the universe.

\section{Cosmography and observable quantities}

In the previous section, we noticed that cosmography is easily obtained by means of a Taylor expansion of the scale factor. We also discussed that such an expansion is not the only possibility, but practically it is the most viable one to define the cosmographic series. More generally, cosmography may even be adapted to any Taylor expansions related to cosmological quantities, by employing the scale factor expansion. This property turns out to give expanded quantities which can be directly matched with observations. As an immediate example, we can consider the pressure and density expansions around $z=0$. Generally speaking, the benefits of doing  this depends on how much the observable quantity, expanded in terms of the cosmographic series, can be matched with data. In other words, although we may be able to get a Taylor series of any quantity in cosmology, the focus should be on those which can be compared directly with data. Hence, the first non-trivial example of cosmographic observable expansion is found by expanding the Hubble rate today \cite{hubblerate1,hubblerate2}.  One can then go further by obtaining the cosmological distances in terms of the cosmographic series \cite{distanza}. Furthermore, cosmography may even enter the energy conditions to produce priors over the violations of such relations as experimentally observed. This matching is an initial point toward a proper cosmographic sound speed definition \cite{ilmiopaperconquew}. From this consideration, one reproduces the sound speed having a new observable fixed by late-time data. In such a picture, one infers a unified dark energy model derived from cosmography, which may correspond to the dark fluid with gaseous contribution \cite{iosuono2}. Finally, a further observable is offered by the equation of state and by its connection to the cosmographic series \cite{equax}. Below, we highlight all those aspects and focus on the role played by the principal distances and their relation with the cosmographic parameters.

\subsection{Cosmography with the Hubble parameter and beyond}

The cosmographic series corresponds to the evaluation of scale factor derivatives at arbitrary orders. An important property appears by rescaling all coefficients in terms of one of them, i.e., the Hubble rate. In fact, by choosing to expand $H$ to order $n$, in terms of the cosmographic series one finds that only $n-1$ cosmographic coefficients are effectively independent. To see this more clearly, we give below the connection between the cosmographic series and the Hubble rate:
\begin{subequations}\label{Hpunto}
\begin{align}
\dot{H} &= -H^2 (1 + q)\,,\\
\ddot{H}  &= H^3 (j + 3q + 2)\,,\\
\dddot{H}  &= H^4 \left [ s - 4j - 3q (q + 4) - 6 \right]\,,\\
\ddddot{H}  &= H^5 \left [ l - 5s + 10 (q + 2) j + 30 (q + 2) q + 24\right ]\,,\\
H^{(5)} &=H^6\left\{m-10j^2 - 120j (q + 1) - 3\Big[2l+5(24q+18q^2+2q^3-2s-qs+8)\Big]\right\}\,.
\end{align}
\end{subequations}
Physically, this choice corresponds to relating the cosmographic coefficients to the Hubble rate, which describes the dynamics of the universe at the present time. More practically, the above formulae points out that the knowledge of $H$ on different scales allows one to reproduce limits on the cosmographic series, giving the corresponding intervals over which all coefficients span. For these reasons, the Hubble rate is sometimes not considered a \emph{genuine} cosmographic parameter, but is rather a cosmological quantity which can be related to the series itself. Furthermore, by using the definition:
\begin{equation}\label{jdf78}
H=\Big[\frac{d}{dz}\left(\frac{d_L}{1+z}\right)\Big]^{-1}\,,
\end{equation}
where $d_L$ is the luminosity distance, defined by\footnote{The ratio of the apparent and the absolute luminosity of an astrophysical object, in terms of the luminosity $L$ and flux $\Phi$.} $d_L=\sqrt{\frac{L}{4\pi \Phi}}$, one obtains:
\begin{equation}
d_L=\frac{z}{H_0}\sum_{n=0}^{N}\alpha_nz^n\,,
\label{dlexpanded10}
\end{equation}
where:
\begin{subequations}
\begin{align}
\alpha_0  & =  1\\
    \alpha_1  &=  \frac{1}{2} - \frac{q_0}{2} \\
    \alpha_2  &=  -\frac{1}{6} -\frac{j_0}{6} + \frac{q_0}{6} + \frac{q_0^2}{2}\\
    \alpha_3  &=  \frac{1}{12} + \frac{5 j_0}{24} - \frac{
		q_0}{12} + \frac{5 j_0 q_0}{12} -    \frac{5 q_0^2}{8} - \frac{5 q_0^3}{8} + \frac{s_0}{24}\\
    \alpha_4  &=  -\frac{1}{20} - \frac{9 j_0}{40} + \frac{j_0^2}{12} - \frac{l_0}{120} +
    \frac{q_0}{20} - \frac{11 j_0 q_0}{12} + \frac{27 q_0^2}{40} - \frac{7 j_0 q_0^2}{8} + \frac{11 q_0^3}{8} +
    \frac{7 q_0^4}{8}\nonumber\\ &- \frac{11 s_0}{120} - \frac{q_0 s_0}{8}\\
    \alpha_5  &=  \frac{1}{30} + \frac{7 j_0}{30} - \frac{19 j_0^2}{72} + \frac{19 l_0}{720} +
    \frac{m_0}{720} - \frac{q_0}{30} + \frac{13 j_0 q_0}{9} - \frac{7 j_0^2 q_0}{18} + \frac{7 l_0 q_0}{240}
    - \frac{7 q_0^2}{10} \nonumber\\ &+ \frac{133 j_0 q_0^2}{48} - \frac{13 q_0^3}{6} +
    +\, \frac{7 j_0 q_0^3}{4} - \frac{133 q_0^4}{48} - \frac{21 q_0^5}{16}+ \frac{13 s_0}{90}
    - \frac{7 j_0 s_0}{144} + \frac{19 q_0 s_0}{48} + \frac{7 q_0^2 s_0}{24}.
\end{align}
\end{subequations}
This expression for $d_L$ is general and applies to any cosmological model. Hence, by directly fitting Eq. \eqref{dlexpanded10} using cosmic data, one gets physical limits on $q_0$, $j_0$, $s_0$ and $l_0$ \emph{for any cosmological model}.

Analogously, it would be interesting to expand commonly used notions of cosmological distances. Following the prescriptions given in~\cite{cattviss2007}, we consider four alternative distances, namely the photon flux distance $d_F$, the photon count distance $d_P$, the deceleration distance $d_Q$ and the angular diameter distance $d_A$. To define those distances in a self-consistent way, we construct the following relations:
\begin{eqnarray}\label{serie2}
    d_L & = & D (1+z) = D \cdot \frac{1}{a(t)}\,, \nonumber \\
    d_F & = & \frac{d_L}{(1+z)^{1/2}} = D \cdot \frac{1}{\sqrt{a(t)}}\,, \nonumber \\
    d_P & = & \frac{d_L}{(1+z)} = D\,, \\
    d_Q & = & \frac{d_L}{(1+z)^{3/2}} = D \cdot \sqrt{a(t)}\,, \nonumber \\
    d_A & = & \frac{d_L}{(1+z)^2} = D\cdot a(t)\,.\nonumber
    \label{eq:dist}
\end{eqnarray}
These measures of distances are less commonly used in cosmology. Apart from the luminosity distance $d_L$, we give here the \emph{photon flux distance}, $d_F$. This is usually not evaluated from the detector energy flux but from the photon flux and consequently turns out to be experimentally much easier to determine. The photon count distance, $d_P$, is essentially built up by the total number of photons which are detected as opposed to the photon rate. The
deceleration distance, $d_Q$, was first introduced by~\cite{cattviss2005}, since it has a useful dependence on the acceleration parameter $q_0$. This is relevant, since $q_0$ is the best-constrained cosmographic parameter, together with $H_0$. It therefore follows that this distance may reveal hints on what priors to use for $H_0$ and $q_0$. The last distance, i.e., the angular diameter distance, $d_A$, was proposed in~\cite{weinberg2008}, as the ratio between the physical size of the object at the time of light emission and its angular diameter observed today.

All those quantities depend on the \emph{physical distance}, i.e., the distance traveled by a photon that is emitted at time $t_1$ and absorbed at our epoch $t_0$, with $t_0>t_1$:
\begin{equation}
D = \int \d t = \Delta t\equiv (t_0 - t_1)\,.
\end{equation}
Under this hypothesis, one might expect $D$ to be useful in observational cosmology to obtain cosmographic constraints. The same hypothesis holds for the other distance definitions, which can be thought as standard cosmographic indicators, as also proposed in \cite{annalen}. In such an approach, one argues the need of using all the distances for a complete cosmographic test, since all the various cosmological distances assume the total photon number is conserved on all cosmic scales. The basic idea is to assume that we are able to obtain constraints from all distances measures and to overlap the obtained bounds in order to reduce the net priors on the cosmographic parameters. In principle, there should be no reason to discard one distance measure in favour of another one, however recently a \emph{duality problem} was found for such distances \cite{hung8}, suggesting that a better understanding of their applicability is required before they can be reliably used in observational cosmology \cite{bll}. Due to this fact, it has been suggested that $d_L$ works better than all other approaches, especially when dealing with supernova data \cite{bll3,bll4}.

\subsection{The cosmographic energy conditions}

In general relativity, the relations related to the energy conditions are not \emph{a priori} known. In cosmology, considering the Friedmann-Robertson-Walker  metric, it is possible to guarantee stability and causality invoking the following relations \cite{carroll}:
\begin{eqnarray}\label{ECs0}
T_{\mu\nu}k^\mu k^\nu\geq0 \quad {\rm (Weak\,\, Energy\,\, Condition)}\;,\\
(T_{\mu\nu}-{1\over2}g_{\mu\nu}T)\phi^\mu\psi^\nu\geq0 \quad {\rm (Strong\,\, Energy\,\, Condition)}\;,\\
T_{\mu\nu}\varphi^\mu \eta^\nu\geq0 \quad {\rm (Dominant\,\, Energy\,\, Condition)}\;,
\end{eqnarray}
where $k^{\mu}$ and $k^\nu$ are null vectors, $\phi^\mu$ and $\phi^\nu$ null time-like vectors whereas $\varphi^\mu$ and $\eta^\nu$ two co-oriented time-like vectors. In the homogeneous and isotropic universe, they read:
\begin{eqnarray}\label{ECs1}
\rho_t &\geq& 0, \quad \rho_t + P \geq 0 \quad {\rm (Weak\,\, Energy\,\, Condition)}\;,\\
\rho_t + 3P &\geq& 0, \quad  \rho_t +P \geq 0 \quad {\rm (Strong\,\, Energy\,\, Condition)}\;,\\
\rho_t - P &\geq& 0, \quad \rho_t +P \geq 0 \quad {\rm (Dominant\,\, Energy\,\, Condition)}\;,
\end{eqnarray}
which are usually referred to as \emph{cosmological energy conditions}. In the cosmographic case, they can be rewritten as follows:
\begin{eqnarray}\label{ECs2}
H_0^2 \geq 0 \quad &{\rm (Weak\,\, Energy\,\, Condition)}\;,\\
-{1\over2}\leq q_0<0  \quad &{\rm (Strong\,\, Energy\,\, Condition)}\;,\\
q_0>2 \quad &{\rm (Dominant\,\, Energy\,\,. Condition)}\;.
\end{eqnarray}
The quite surprising outcome is that only the weak energy condition trivially holds if $H$ is real, while all the others are \emph{strongly} violated, in order for the universe to have an accelerated expansion. Combining theoretical expectations with the energy conditions suggests instead:
\begin{equation}\label{ECs3}
H_0>0\,,\qquad -1<q_0\leq-\frac{1}{2}\,.
\end{equation}
The energy conditions therefore lead to cosmographic limits on the Ricci scalar. This may be viewed as a first approach to constrain the Ricci scalar itself, which in the Friedmann-Robertson-Walker metric is given by
\begin{equation}
R=-6\left(\dot H+2H^2\right)=-\rho+3P\;.
\end{equation}
In this way, we can obtain bounds on $R$ at the present time, i.e., $R_0$, and we can give limits on its variation in time. For example, the first two terms are:
\begin{subequations}\label{ricci12}
\begin{align}
R_0&=-6H_0^2(q_0-1)\,,\\
\dot R_0&=-6H_0^3(j_0+2q_0+1)\,.
\end{align}
\end{subequations}

\subsection{The cosmographic sound speed - A way to obtain a unified dark energy model}

Under the assumption of a of perfect fluid, the adiabatic sound speed $c_s$ becomes a free parameter, sometimes considered as an additional observable, since it can be used in both late and early-time cosmology. In particular, the clustering of dark energy is more efficient when over-densities are not balanced by local pressure perturbations\footnote{The case of canonically kinetic scalar field is quite different. Even if it fluctuates, the sound speed is $c_s=1$, but the local pressure prevents the density contrast from growing significantly.}. Following the standard definition \cite{defsound}, we get:
\begin{equation}\label{eqsound}
c_s^2 \equiv \left(\frac{\partial P}{\partial \rho}\right)_S\,,
\end{equation}
taking the entropy $S$ to be constant. The sound speed enters the Jeans length $\lambda_J$ and so depending on its sign, one obtains limits on structure formation on different scales. A vanishing sound speed corresponds to perturbations growing on all scales. The \emph{cosmographic sound speed} can be easily determined:
\begin{equation}\label{eqsound}
c_s^2 = \left(\frac{\partial P}{\partial \rho}\right)_S = \frac{1}{3}\frac{j-1}{1+q}\,,
\end{equation}
which simultaneously implies $j\geq1$ and $q>-1$ or $j<1$ and $q<-1$ in order to guarantee that $c_s>0$. These limits satisfy Eq. \eqref{ECs3} only for the case: $j\geq1$ and $q>-1$. As the sound speed vanishes\footnote{If one considers a particular case in which a polynomial scale factor reproduces the sound speed in the observable regime, then one notices that only the case $n\geq\frac{2}{3}$ predicts a physical sound speed, which vanishes exactly when $n=\frac{2}{3}$. In such a picture, it is possible to exclude $n<0$ which corresponds to singular scale factor. Finally, at $n=0$ and $n=\frac{1}{2}$ the sound speed provides  a divergence \cite{ellyx}.}, which corresponds to having fluid perturbations growing on all scales\footnote{As cold dark matter appears to do, in agreement with observations \cite{darkma}.}, it is also possible to get different manifestations of the same fluid. The corresponding paradigm has been preliminary investigated in \cite{ilmiopaperconquew,iosuono2,suono3} and represents a unified dark energy model, usually named \emph{dark fluid}~\cite{Hu:1998tj}. The model assumes that the pressure is constant with an evolving barotropic factor, differently from $\Lambda$CDM in which the density, pressure and the barotropic factor are constant. In this case, assuming the pressure to be constant, it is possible to demonstrate that the barotropic factor and density are functions of the redshift\footnote{The debate on the ability to distinguish $\Lambda$CDM from a dark fluid is still open \cite{debate}.}, since $\omega\sim\rho^{-1}$.

The first and probably easiest physical application of such an approach deals with assuming a dark energy in the form of ideal gas with zero sound speed at all scales as firstly noticed in \cite{ilmiopaperconquew}. Integrating the density, one gets:
\begin{equation} \label{DarkFluidRev:rhodEv}
 \rho_d(a) = \frac{\rho_{d\,0}}{1+ \mathcal{K}} \left( 1 +   \frac{\mathcal{K}}{a^3} \right)\,,
\end{equation}
with the constant $\mathcal{K} = - (\rho_{d0} + P_0)/P_0$, where $\rho_{d0}$ is the dark fluid energy density at $a(t_0) \equiv a_0 = 1$. The corresponding model unifies both dark energy and matter under a single framework. In other words, in order to build the cosmological model, we need to consider only baryons and dark energy contributions, with dark matter arising as a consequence of the model itself. Other kinds of unified dark energy model generally do not provide a vanishing sound speed \cite{nosoundzero,nojiri} and so, since the $\Lambda$CDM model has a vanishing sound speed, it is evident that the dark fluid picture degenerates with the minimal model. This degeneration is not removed at a fundamental level since, from comparisons with the observed large scale matter power spectrum, the model is effectively indistinguishable from $\Lambda$CDM \cite{Aviles:2011ak}.

From Eq.~(\ref{DarkFluidRev:rhodEv}), one notices a component that decays $\propto a^3$ plus a component that remains constant at different stages of the universe's evolution. The constant $\mathcal{K}$ is assumed positive to enable acceleration today, showing a corresponding pressure in the interval $-\rho_{d0}\le P_0 \le 0$, with a barotropic equation of state given by:
\begin{equation} \label{DarkFluidRev:DFEoS}
 w_d(a) = -\frac{1}{1+\mathcal{K}a^{-3}}\,.
\end{equation}
In comparison with the $\Lambda$CDM model, one gets
\begin{equation} \label{DarkFluidRev:LDMEoS}
 w_{\Lambda + dm }(a) =  -\frac{1}{1+\frac{\Omega_{dm}}{\Omega_\Lambda}a^{-3}}\,,
\end{equation}
which when compared gives:
\begin{equation} \label{DarkFluidRev:DFLCDMId}
 \mathcal{K} = \frac{\Omega_{dm}}{\Omega_\Lambda}\,,
 \qquad \rm{and} \qquad \Omega_{d} = \Omega_{dm} + \Omega_\Lambda\,.
\end{equation}
It is easy to show that the suggestive form $P_0= - \rho_c \Omega_\Lambda$, which is what one expects even in the $\Lambda$CDM case. Although no significant differences arise between the two models\footnote{Heuristic arguments point out that degeneracy is not removed at all orders in perturbation theory.}, we note an remarkable physical difference: when one employs Eq.~(\ref{DarkFluidRev:LDMEoS}) it is strictly necessary to add the cosmological constant into Einstein's energy momentum tensor. On the contrary, in Eq.~(\ref{DarkFluidRev:DFEoS}) the cosmological constant is not strictly necessary. Even though the model may be determined by fixing the cosmographic coefficients, in particular, for $q$ and $j$, the cosmographic approach by itself is unable to remove the degeneracy. This is due to the fact that the cosmographic series, or alternatively speaking the equation of state, are quantities which account for the total energy density and pressure. If one could somehow fix limits on the dark energy cosmographic coefficients only, it would be possible to distinguish the two approaches and to remove the degeneracy.

\subsection{The effective thermodynamic approach in cosmography}

Throughout the text we have assumed that the gravitational source is modelled by perfect fluids. Consequently, in the arena of cosmography, one can describe the universe as a thermodynamic system in which all thermodynamic quantities are related to the cosmographic series \cite{pad}. This is supported by the fact that  the laws of thermodynamics are mathematically consistent with the Friedmann-Robertson-Walker metric \cite{herny} and consequently the first law of thermodynamics leads to a given expression for the temperature of the universe. Analogously, following standard definitions, one can determine the thermodynamic quantities in the context of a homogeneous and isotropic universe, where
\begin{equation}\label{sph}
\textsf{C}\equiv\frac{\partial \hat U}{\partial T}
\end{equation}
and
\begin{equation}\label{sph}
\textsf{S}\equiv V\frac{P+\rho}{T}\,,
\end{equation}
are the specific heat and entropy respectively. Assuming that the function $\hat U$ is the internal energy $U$ and the enthalpy $h$, we obtain
\begin{subequations}\label{calorrr}
\begin{align}
\textsf{C}_V &= \left(\frac{\partial U}{\partial T}\right)_V\,,\\
\textsf{C}_P &= \left(\frac{\partial h}{\partial T}\right)_P\,,
\end{align}
\end{subequations}
which gives the specific heat at constant pressure and volume respectively. Moreover, since $U^\prime = \left(\frac{\partial U}{\partial T}\right)_V\,T^\prime + \left(\frac{\partial U}{\partial T}\right)_T\,V^\prime$ and $h^\prime = \left(\frac{\partial h}{\partial T}\right)_P\,T^\prime + \left(\frac{\partial h}{\partial P}\right)_T\,P^\prime$, we have from Eq. \eqref{calorrr}

\begin{subequations}\label{calorrr2}
\begin{align}
\textsf{C}_V &= \frac{1}{T^\prime}\Big[h^\prime - \left(\frac{\partial U}{\partial T}\right)_V\left(\frac{\partial P}{\partial S}\right)_V V^\prime-VP^\prime\Big]\,,\\
\textsf{C}_P &= \textsf{C}_V + \frac{1}{T^\prime}\Big[\left(\frac{\partial U}{\partial T}\right)_V\left(\frac{\partial P}{\partial S}\right)_V V^\prime+VP^\prime-\left(\frac{\partial h}{\partial P}\right)_T P^\prime\Big]\,.
\end{align}
\end{subequations}
The specific heats behave differently depending on the choice of the underlying cosmological model. In the simplest case of a gaseous dark energy contribution, in terms of the cosmographic series, one obtains
\begin{subequations}\label{bouh}
\begin{align}
\textsf{C}_P&=\frac{V_0}{4\pi G}\frac{H^2}{T^{'}}\frac{j-1}{(1+z)^4}\,,\\
\textsf{C}_V&=\frac{V_0}{8\pi G}\frac{H^2}{T^{'}}\frac{2q-1}{(1+z)^4}\,,\nonumber
\end{align}
\end{subequations}
where we made use of $U= V_0\rho a^{3}$ and $h = V_0\left(\rho+P\right)a^3$. It follows that $U$ and $h$ are positive quantities, whereas the sign of $\textsf{C}_{P0}$ is a function of $j_0-1$, giving $j_0>0$, $\textsf{C}_P>0$, and $\textsf{C}_V<0$, since $q_0<0$\footnote{The temperature variation with respect to the redshift $z$ decreases as the universe expands, so that $T^\prime>0$.}. The entropy in terms of the cosmographic series reads
\begin{equation}\label{entropia}
S=\frac{2H^2q}{T}\;.
\end{equation}
In the same way, one gets the enthalpy as:
\begin{equation}\label{ental}
h=(\rho+P)V=ST=2H^2q\,.
\end{equation}
These quantities are examples of measurable thermodynamic quantities in terms of the cosmographic series. In fact, in our picture, since $H^2$ and $T$ are positive definite, we demonstrate that both $S$ and $h$ should be negative. This hypothesis underlines the fact that dark energy is supposed to evolve as a gas. In cases where this is not true, the Mayer relation does not necessarily have to hold and the specific heats become more complicated. Consequently the thermodynamic quantities are not forced to be negative\footnote{At the background level, it could happen that $C_P = C_V$, when the pressure is constant, discarding the condition of perfect gas \cite{ioavileseklapp}.}.

\subsection{From cosmography to the equation of state - What next?}

We discussed in Sec. II that the equation of state of the universe can be defined as the ratio between the total pressure and the total density constituting the whole energy budget, i.e., $\omega\equiv\frac{\sum_iP_i}{\sum_i\rho_i}$. A formal difference is present if one considers the equation of state of a single component; in this case we have: $\omega_i\equiv\frac{P_i}{\rho_i}$, with a specified index $i$, corresponding to e.g.,  matter, curvature, radiation, neutrinos, and so forth. The two expressions differ from each other significantly. In general, it is hard to disentangle $\omega_i$ and $\omega$, measuring simultaneously both these terms. The problem is present both in the background and in perturbation theory. As we stressed above, cosmography only provides constraints from the second Friedmann equation giving: $\omega = \frac{P}{\rho} = \frac{2q-1}{3}$, and so:
\begin{subequations}\label{omegazvarie}
\begin{align}
\frac{\d \omega}{\d z}&=\frac{j-2q^2-q}{1+z}\,,\\
\frac{\d^2 \omega}{\d z^2}&=-\frac{q(j-2)-8q^2(1+q)+s}{(1+z)^2}\\
\frac{\d^3 \omega}{\d z^3}&=\frac{l-j^2-6q(1+2q)^3+j[-14+q(-7+23q)]+5(1+q)s}{(1+z)^2}\,,\\
\frac{\d^4 \omega}{\d z^4}&=\frac{-m+24q(1+2q)^4-8j^2(1-5q)-2l(6+5q)-(16+q(48+53q))s}{(1+z)^4}\nonumber\\
&\quad +\frac{j(88+q(60-q(232 + 215 q))+7s)}{(1+z)^4}\,.
\end{align}
\end{subequations}
For single species, we can compute the cosmographic series by hand. For example, knowing that $\omega_m=0\,,~\omega_k=-\frac{2}{3}\,,~ \omega_r = \frac{1}{3}$ respectively for standard pressureless matter (baryons and dark matter), spatial curvature and radiation, we have $\omega_{DE}=\frac{ P_{DE} }{\rho-\rho_m}$, where we have neglected radiation and spatial curvature and clearly $\omega_{DE}\neq\omega$. Hence, the corresponding cosmographic series takes different set of values depending on whether the universe is built up by from one, two or more fluids, having $\rho>\rho_{DE}$. This degeneracy jeopardises the standard approach and can be recast as \cite{rubano}:
\begin{equation}\label{recadege}
\omega\equiv\frac{\sum_iP_i}{\sum_i\rho_i}=-\frac{1}{3}\left(\frac{2\dot{H}+3H^2}{H^2-H_{0}^{2}\Omega_{m,0}a^{-3}}\right)\,.
\end{equation}
For barotropic fluids, the cosmological model is determined once $P(\rho)$ is known. Alternatively, one needs the corresponding equation of state for a particular species, i.e., $\omega_i$, to obtain the corresponding evolution in terms of the redshift. We can easily obtain different outcomes coming from postulating the equation of state for single species. For example, $q_0=0$ in the case of matter, radiation, spatial curvature equations of state only, whereas $q_0=-1$ for the cosmological constant equation of state. Analogously, $j_0=1$ for all those cases respectively, while $s_0=2$, except for the case of the cosmological constant in which $s_0=1$.

\section{Cosmography and cosmological models}

One of the most important advantages of cosmography is that it may be used to determine constraints on cosmological models in two distinct ways, i.e., the \emph{direct and inverse methods}. The first strategy consists of relating the free parameters of a given model to the cosmographic series. In this way, fixing limits on the cosmographic series corresponds to setting the intervals over which the free parameters span\footnote{This can be also used to define priors on the free constants of a given model.}.

In particular, assuming that a given model depends on a vector of free parameters of the form: $\xi_{i}$ with $i=1,\ldots, N$, we can formally write:
\begin{eqnarray}\label{basis}
q_0=q_0\left(\xi_i\right) \,,\qquad j_0=j_0\left(\xi_i\right) \,,\qquad s_0=s_0\left(\xi_i\right)\qquad l_0=l_0\left(\xi_i\right)\,,\qquad m_0=m_0\left(\xi_i\right)\,.\nonumber
\end{eqnarray}
Hence, since the luminosity curves are one-to-one invertible, one obtains:
\begin{eqnarray}\label{inbasis}
\xi_i=\xi_i\left(q_0,j_0,s_0,l_0,m_0\right)\,.
\end{eqnarray}
In other words, once the cosmographic series is measured, the vector $\xi_i$ is uniquely determined. However, the disadvantage of the direct method is the fact that the corresponding errors on $\xi_i$ are typically large, since they propagate through the logarithmic rule. In fact, discarding systematics and assuming $\Psi\in\xi_i$ to be a generic parameter of a certain model, we get $\Psi=\Psi(\alpha_{1},\ldots,\alpha_{N})$, with $\alpha_{i}$, $\forall i\in[1,\ldots,N]$. Thus, we obtain:
\begin{equation}\label{g}
\delta\Psi=\sum_{i}^{N}\Big|\frac{\partial \Psi}{\partial
\alpha_{i}}\Big|\times\delta\alpha_{i} \;,
\end{equation}
giving errors associated to the $i$-th variable: $\delta\alpha_i$ and so $\Psi\pm\delta\Psi$. For the sake of clearness, Eq. \eqref{g} represents a lower limit on error estimates, since a complete description of total errors on $\Psi$ depends also on systematics $\sigma_s$. A way of including systematics could be achieved by assuming \cite{lib12}
\begin{equation}\label{g1}
\delta\Psi_{tot}=\sqrt{(\delta\psi)^2+\sigma_s^2}\,.
\end{equation}
Over-optimistic estimates of systematic uncertainties does not modify the cosmographic analyses, but instead reduces the significance on the constraints on $\Psi$.

The inverse procedure significantly improves the above results. Assuming, in fact, the validity of Eq. \eqref{basis}, one can imagine replacing the cosmographic coefficients in terms of $\xi_i$. In this way, the observable quantities adopted for the cosmographic fits are \emph{directly} measured by means of a new sets of ''cosmographic coefficients''. To understand this, imagine using $d_L$ to make a cosmographic measurement. We find that there exists a direct correspondence $d_L=d_L(H_0,q_0,j_0,s_0,\ldots)$ and so the set $H_0,q_0,j_0,s_0,\ldots$ is uniquely determined. Now, if we consider instead the set $\xi_i$ to rewrite $d_L$ as $d_L=d_L(\xi_i)$, the luminosity distance depends upon $\Psi_j\in\xi_i, \forall j, \forall i$. In this way, one does not have to statistically propagate the errors, which are simply the ones obtained from the analyses itself. The corresponding errors are assumed to be smaller than the ones obtained in the direct procedure.

\subsection{A generic cosmographic model}

The requirement for building up a cosmological model which is able to fit the cosmographic prescription at small redshift is an object of debate. The problem is most likely due to the degeneracy which jeopardises the cosmographic approach at late-times \cite{deg88}.

However, every evolving dark energy paradigm should satisfy some cosmographic recipes, which may impose stringent limitations on the dark energy expansion history. To better understand this fact, let us consider a generic Hubble rate in which the evolution of dark energy is thought to be given by the form of a unknown function $G(z)$. Hence, we simply have:
\begin{equation}
H = H_0 \sqrt{\Omega_m (1 + z)^3 + G(z)}\,.
\end{equation}
Simple calculations give us\footnote{In principle, it would be possible to use all the cosmographic parameters, however, here we report on only two of them: $q_0, j_0$. We made this choice because of the large expressions obtained when going to higher order in the cosmographic expansions and because one does not need further orders in order to recover the dark energy behavior at small redshifts.}:
\begin{subequations}\label{dinamicamente1}
\begin{align}
q_0&=-1 + \frac{3 \Omega_m + G'(0)}{2 (\Omega_m + G(0))}\,,\\
j_0&= \frac{ \Omega_m +  G(0) -  G'(0)+ G''(0)/2}{\Omega_m + G(0)}\,.
\end{align}
\end{subequations}
By inverting, we get
\begin{subequations}\label{dinamicamente2}
\begin{align}
G'(0)&= \Omega_m(2q_0-1)+2 G(0)(q_0+1)\,,\\
G''(0)&=2\Big\{\Omega_m[j_0 + 2(q_0-1)] + (1 + j_0 + 2 q_0) G(0)\Big\}\;.
\end{align}
\end{subequations}
The expressions of $G^\prime(0),G^{''}(0)$ are evaluated once the cosmographic series is known from numerical fits, using the assumption that $G(0)=1-\Omega_m$. The opposite view consists of reproducing viable priors on the cosmographic series by investigating the phase space of $G^\prime(0),G^{''}(0)$. From that space one may obtain allowed regions in which the cosmographic coefficients span\footnote{The first approach is essentially due to the direct procedure whereas the second treatment is based on the inverse procedure.}.

In \eqref{dinamicamente2} the obtained viable examples of $G(0), G^\prime(0), G^{''}(0)$ from the intervals of $q_0, j_0$, with $\Omega_m=0.315^{+0.013}_{-0.013}$, $q_0=0.561^{+0.055}_{-0.042}$ and $q_0=0.999^{+0.346}_{-0.468}$ \cite{ade,gruberluongo} are:
\begin{subequations}\label{numerante}
\begin{align}
G(0)&\in[0.672\,;\,0.698]\,,\\
G'(0)&\in[-0.112\,;\,0.004]\,,\\
G''(0)&\in[-1.162\,;0.698]\,.
\end{align}
\end{subequations}
There exist several ways to reconstruct the dark energy equation of state \cite{ricostruzione1,ricostruzione2}. Every method, however, suffers from theoretical issues and so the need of having a hint of how dark energy evolves is essential in order to determine possible departures from the Concordance model. One possible assumption is that the effects of $\Lambda$CDM are a limiting case of cosmology at the present time.
In addition, one can demand that:
\begin{itemize}
  \item the dark energy evolution is slightly variable with respect to the redshift $z$, i.e., it has to weakly increase for increasing redshift;
  \item the dark energy model must not admit divergences either at high redshift or at future times, in order to be compatible with observations of the matter power spectrum;
  \item the model should be predictive concerning the values of the pressure and density today and in intermediate phases of the universe's evolution. It must also be compatible with observations of the transition redshift, at which dark energy becomes dominant over matter.
\end{itemize}
To guarantee that these conditions hold, a simple possibility is to assume a dark energy term $\Omega_X$ of the following form:
\begin{equation}
\Omega_{X} = A\cdot \Bigg[1 +\frac{\log(\alpha + \beta z)}{(1 + z)}\Bigg]\,,
\end{equation}
where $\alpha$ is not a cosmographic parameter since
\begin{equation}
\alpha= e^{\Bigl(\frac{1 - \Omega_m}{A} - 1\Bigr)}\,.
\end{equation}
A quantitative analysis is presented in the figure below, Fig. \ref{vedisotto}

\begin{figure}[ht]
\includegraphics[width=4in]{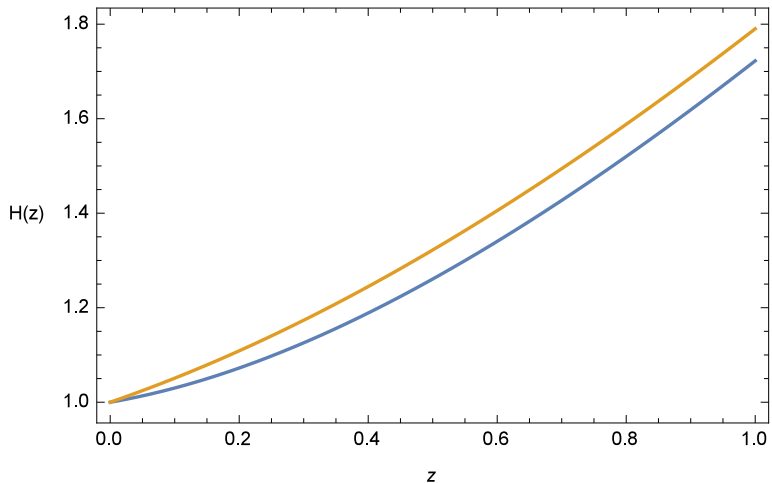}
\caption{Plot of the functional behavior of $H(z)$ up to redshift $z=1$. We considered the indicative values $A=1$ and $\beta=0.5$.}
\label{vedisotto}
\end{figure}

\subsection{The $\Lambda$CDM model}

The modern \emph{minimalist model} which is now considered as the Concordance paradigm describes the expansion history of the universe by using a constant dark energy term\footnote{For the sake of clearness, it is slightly misleading to refer to a \emph{constant dark energy} term. In fact, the dark energy contribution has a physical interpretation which is completely different from the one connected to the cosmological constant. The cosmological constant, in fact, is related to pure vacuum energy, while dark energy is supposed to evolve during the entire expansion history of the universe in a different manner. However, we believe that naively one can refer to the cosmological constant as a sort of constant dark energy term.}. The corresponding paradigm, named the $\Lambda$CDM model, consists of a cosmological constant $\Lambda$ and pressure-less matter, with vanishing spatial curvature. In particular, the matter contribution is \emph{assumed to be composed} of dark matter and baryons\footnote{An important fact is that this model and even a wide number of other paradigms absorb dark matter into the total matter density. In other words, the mass density includes both dark matter and baryons in agreement with observations. In unified dark energy models, however, the cold dark matter contribution comes from dark energy through an emergent mechanism, without assuming it \emph{a priori} to be included in the energy momentum tensor.}. The original origin of $\Lambda$ arose from the philosophical hypothesis that the universe is static\footnote{This hypothesis was proposed by Einstein himself and after the discovery of cosmic expansion was ruled out.}. Hence, the original meaning of this term was very different from the present one. Notwithstanding, its re-introduction is able to explain the acceleration of the universe, by simply making the shift:
\begin{eqnarray}\label{perho}
\rho \rightarrow \rho + \frac{\Lambda}{8\pi G}\,,\qquad
P \rightarrow P -\frac{\Lambda}{8\pi G}\,,
\end{eqnarray}
which leaves unaltered the Friedmann equations.

The advantage of the concordance model lies in its simplicity\footnote{The model depends on one parameter only, i.e., $\Omega_m$, if $H_0$ is fixed or marginalised.} and on its ability to describe the universe's expansion history. Furthermore, it seems to be able to reproduce both the late and early-phases of the evolution of the universe. Despite these successes, the model is plagued by two major issues\footnote{It is not clear whether the model is plagued by \emph{only} two (theoretical) drawbacks or whether there is something beyond. There is a consensus around the belief that the model \emph{should be} the definitive one, but also a criticism around the introduction of $\Lambda$ itself, which comes from other physical issues which appear to plague the model \cite{ornot}.}. Theoretically, it seems difficult to explain why we live in a precise epoch, in which the cosmological constant density is of the same order of magnitude as the  matter density\footnote{The ratio between the two quantities is $\sim3$ in favour of the cosmological constant.}. This is known as the \emph{coincidence problem}. On the other side, there exists also a huge fine-tuning problem between the measured and predicted values of the energy density of $\Lambda$, which is of the order of $10^{120}$ orders of magnitude larger than the observational value \cite{ilmodello}.

These two theoretical ''catastrophes'' may suggest that beyond the current concordance model, an evolving dark energy reduced to a cosmological constant at small redshifts. The $\Lambda$CDM Hubble rate is given by
\begin{equation}\label{nncivo}
H=H_0\sqrt{\Omega_{m}(1+z)^3+\Omega_k(1+z)^2+\Omega_\Lambda} \,
\end{equation}
with $\Omega_\Lambda=1-\Omega_m-\Omega_k$. The cosmographic series is given in the Appendix.

\subsection{The $\omega$CDM model}

The simplest model which extends the $\Lambda$CDM paradigm is the $\omega$CDM approach. This paradigm assumes a constant barotropic factor, hereafter $\omega$, which departures from the $\Lambda$CDM model, since it does not provide \emph{exactly} $\omega\neq-1$. In analogy to the concordance model, in order to guarantee that the universe accelerates, one needs a negative $\omega$, with the condition\footnote{However, observations do not univocally demonstrate that the case $\omega<-1$ should be ruled out.} $\omega>-1$. To physically motivate this model, one can consider a \emph{quintessence fluid} described by means of a \mbox{slowly-rolling} dynamical scalar field, with a non-canonical kinetic term in the Lagrangian.

Hence, the Hubble parameter, having fixed $H_0$, is dependent on two parameters and it becomes slightly more complicated than the standard Concordance paradigm, since dark energy evolves in time. We therefore have:
\begin{equation}\label{XCDM}
H=H_0\sqrt{\Omega_{m}(1+z)^3+\Omega_k(1+z)^2+(1-\Omega_m-\Omega_k)(1+z)^{3(\omega+1)}}\,.
\end{equation}
The free parameters are $\Omega_m$ and $\omega$. The corresponding cosmographic series is given in the Appendix.

\subsection{The Chevallier-Polarski-Linder parametrization}

There exists a wide consensus that it is advantageous to express the barotropic factor in terms of the cosmic time or more properly in function of either the scale factor or the redshift $z$. This property is physically motivated by the fact that in principle dark energy may evolve throughout the history of the universe via some generic (still to be determined) function. The simplest hypothesis that one could imagine is to expand $\omega$ as a Taylor series in the redshift $z$:
\begin{equation}\label{expax}
\omega(z)=\sum_{n=0}^{\infty} \omega_n z^n\,.
\end{equation}
However, proceeding with Eq. \eqref{expax} may give misleading results since, at arbitrary order $n$, the case $z\rightarrow\infty$ leads to a divergence in the equation of state\footnote{This is in part due to the fact that the expansion of of $\omega$ has been obtained around $z=0$. Hence, this series is not likely to be valid when $z\rightarrow\infty$.}.

A reasonable approach, first proposed by Chevallier-Polarski \cite{cp} and immediately afterwards by Linder \cite{elle}, with two other independent researchers, pointed out that a direct generalisation of Eq. \eqref{expax} may be\footnote{The model is known in the literature as the CPL parametrisation.}
\begin{equation}\label{expax2}
\omega(z)=\sum_{n=0}^{\infty} \omega_n (1-a)^n\,,
\end{equation}
with $a=\frac{1}{1+z}$. The choice of $\propto 1-a$ instead of $\propto z$ becomes effectively more physical since at the small perturbation regime, no divergences occur. The basic requirement of the CPL parametrisation is to truncate Eq. (\ref{expax}) at first order in $a(t)$. So that, one has $\omega=\omega_0+\omega_1(1-a)$. Here, the Hubble rate is given by:
\begin{equation}\label{cpl2}
H=H_0\sqrt{\Omega_{m}(1+z)+\Omega_k(1+z)^2+(1-\Omega_m-\Omega_k)(1+z)^{3(1+\omega_0+\omega_1)}\exp\left(-\frac{3\omega_1
z}{1+z}\right)} \,.
\end{equation}
The disadvantage is that, once $H_0$ is fixed, the model depends on three parameters: $\Omega_m,\omega_0,\omega_1$, that are not known \emph{a priori}. This complication introduces a further parameter, $\omega_1$, which is weakly constrained by observations. The cosmographic series for this case is given in the Appendix.

\section{The limits and principal drawbacks of cosmography}

In the previous sections, we learned that cosmography represents an extremely accurate technique for investigating the universe's  expansion history in terms of model-independent quantities. Furthermore, we demonstrated that in principle cosmography is also able to directly find bounds on the Taylor derivatives using cosmological data, since all cosmographic derivatives are connected to each other. As a consequence, cosmography does not need to impose \emph{a priori} any cosmological model. However, the cosmographic approach, while powerful and straightforward to implement, is plagued by several shortcomings which either limit its use or forces cosmography to be non-predictive in several cases. The problem is essentially due to the fact that cosmological data, used to fix the cosmographic constraints, are not enough to correctly guarantee that cosmography works well at earlier stages of universe's evolution. In principle, the knowledge of either an infinite number of combined data sets or {\em mock data}, with a large amount of points would better fix the cosmographic parameters, with lower errors. To the best of our knowledge, cosmography might also fail to be predictive due to the fact that it is based on expanded Taylor series, at arbitrary truncated orders. In what follows, we present a list of the most important limits of cosmography, proposing for each of them possible solutions or alternative methods for obtaining (future) more viable cosmographic outcomes.

\subsection{The degeneracy problem among cosmographic coefficients}

The degeneracy problem is a serious drawback which plagues cosmology in general. It is due to the fact that several classes of cosmological models are compatible with cosmological data at small redshift. It follows that the reconstruction of the expansion history of the universe may be obtained by completely different cosmological models, without being able to distinguish between them. This reflects the fact that every paradigm, in turn, is able to describe observations with almost the same statistical precision. Even though coarse-grained methods to statistically discriminate between various models exist, none of them are able to predict which dark energy fluid is favoured and whether the concordance model is the best paradigm or not. Usually, the total errors on measurements are adjusted in a way which assumes that the paradigm under investigation is the favoured one to fit the data. From this assumption, it follows that parameter inference and error bars obtained by taking into account present data with our statistical analyses are effectively unable to investigate problems of pure \emph{model selection}. The situation in cosmography is affected by by analogous problems. However, at a first glance one can think that cosmography is able to get around such caveats by fixing numerical bounds on the cosmographic series itself\footnote{Basically, two separate situations may occur:\\
$1)$ the universe is built up in terms of a single fluid, namely the dark fluid. We showed that here dark matter is unified to dark energy since it emerges as a consequence of assuming one fluid only;
$2)$ the universe's energy momentum tensor is decomposed into (at least) dark energy and dark matter. These two distinct fluids do not interact with each other and evolve differently. If one somehow gets well-constrained cosmographic coefficients at arbitrary redshift regimes, it would be possible to understand whether the universe behaves as a single fluid or by means of a sum of differently evolving constituents. As quoted, however, current data are not enough to focus on the differences between the two approaches.}.
It is expected that robust evidence leading to the understanding of the evolution of dark energy may be obtained by investigating $j$ and $s$. Putting bounds on these terms would enable one to understand at which time the acceleration took place and how dark energy evolves. The degeneracy problem in cosmology would be therefore alleviated if the cosmographic coefficients are better constrained through observations. Unfortunately, when one assumes a particular cosmological distance, there is a degeneracy between all of the cosmographic coefficients, since one cannot measure each of them separately but only the sum. In other words, it is impossible to \emph{disentangle} the cosmographic parameters and so, depending on the probability distribution associated with each coefficient, one would get different results. This issue leads to a degeneracy between cosmographic coefficients and as a consequence the degeneracy between cosmological models remains (since cosmography is unable to remove it with current data sets). Furthermore, the Hubble rate today exhibits a further degeneracy with all the other coefficients, since at first order in the Taylor expansion, every distance measure reduces to $d_i\sim\frac{1}{H_0}z$. This leads to a multiplicative degeneracy among $H_0$ and all the other terms.

A naive way to remove such a degeneration is to marginalise over $H_0$ during a statistical analysis. In such a picture, one is forced to assume multivariate distributions, with probability distribution given for example by: $P(\bf{x},\bf{y})$. Hence, by splitting the cosmographic series as $\vartheta = \Big(H_0,X\Big)$, where $\bf{x}\equiv X$ is the set of all the cosmographic coefficients, except $\bf{y}\equiv H_0$, we can infer the probability distribution of $X$ regardless of the values of $H_0$ by:
\begin{equation}\label{marginalizzazione}
P(X)=\int dH_0 P(H_0,X)\,.
\end{equation}
Alternatively, two further techniques are allowed. The first consists of fitting $H_0$ alone, truncating the Taylor expansion at the lowest order possible. For example, assuming the lowest order of $d_L$, having $d_L\sim \frac{z}{H_0}$, it is possible to get bounds with cosmic data within shorter redshift intervals\footnote{A viable interval would be $z\leq0.5$, in which the union 2.1 data are deeply concentrated.}, in which one can recover an acceptable value for $H_0$ to use in any cosmographic analyses, performed at arbitrary expansion order\footnote{In general, this procedure consists in fixing a coefficient, instead of marginalising over it. Generally speaking, this is not exactly the most practically way to circumscribe the problem, from a pure statistical point of view. However, no significative departures are actually expected once $H_0$ is known from this measurement and so it is possible to allow this technique inside cosmography.}. The second possibility is to fix $H_0$ by means of other types of measurements, either through a $\chi$ square analysis or by fixing it. In all cases, the only possibility of improving the quality of $H_0$ measurements, simultaneously with all the other cosmographic coefficients seems to be to combine different data sets. This provides a way of restricting the parameter space, thus getting more stringent cosmographic results. Summarising, all those considerations suggest that we cannot consider $H_0$ as a \emph{pure} cosmographic parameter, but more precisely as some sort of  \emph{initial condition} to calibrate the cosmic curves. Depending on its value, it could happen that the other terms are well or badly constrained. Thus, the need of such a calibration becomes essential for cosmography and should not be considered irrelevant in order to obtain constraints.

\subsection{The degeneracy with spatial curvature}

Another grave type of degeneracy is due to the spatial curvature. Its role, in the cosmographic constraints, is far from trivial, since $\Omega_k \neq 0$ induces a time variation for the dark energy equation of state \cite{chris2007}, consequently fixing the form of dark energy \emph{a priori}\footnote{Using cosmography to understand if the cosmological model provides an evolving dark energy term should be independent on assuming that the dark energy evolves with time. This fact should be a consequence of the cosmographic recipe, rather than a postulate, assuming $\Omega_k\neq0$.}. Unfortunately, spatial curvature is intimately related to the luminosity distance, since the Friedmann-Robertson-Walker metric formally depends on its value, which is not known \emph{a priori} without observations. In doing so, one is forced to fix $\Omega_k$ to determine limits on cosmography, consequently getting a fixed-curvature cosmographic series. Moreover, the consequence of introducing $\Omega_k$ is that photon geodesic motion changes and this causes a severe degeneration among the cosmographic coefficients and $\Omega_k$ itself. For example, truncating at third order, $j$ cannot be measured alone if $\Omega_k$ is not fixed to a precise value. On the one hand, although precise measurements\footnote{These measurements confirm that $\Omega_k$ may be compatible with zero, but not necessarily perfectly vanishing.} suggest that $\Omega_k\sim 0$, the assumption of  $\Omega_k=0$ imposes a \emph{flat prior} which may influence the reconstructions of dark energy models. On the other hand, if one does not postulate the value of $\Omega_k$ in the cosmographic series, it could happen that a bad convergence arises when constraining the cosmographic parameters. A major challenge, instead of fixing or imposing flat priors with different values of $\Omega_k$, is to get alternative forms of distance, which inform us of the shape of the cosmographic curve regardless of the spatial curvature $\Omega_k$.

It is easy to show that the physical distance $D$, introduced above, alleviates the problem, since it does not depend upon $\Omega_k$. In fact, we obtain\footnote{For brevity, we here report only a fourth order Taylor expansion. Later on, in the appendix, we give the higher order terms.}:
\begin{eqnarray}
D = {\; z\over H_0}
\Bigg\{ 1
-
\left[1+{q_0\over2}\right] {z}
+
\left[ 1 + q_0 + {q_0^2\over2} - {j_0\over6}   \right] z^2
\\
-
\left[ 1 +{3\over2}q_0(1+q_0)+{5\over8}q_0^3-{1\over2}j_0
- {5\over12} q_0 j_0 -{s_0\over24} \right] z^3
+  O(z^4) \Bigg\}\,,
\nonumber
\label{E:physical}
\end{eqnarray}
However, the problem of how to use $D$, instead of $d_L$, still persists since we do not known if $D$ is perfectly related to cosmic data. In other words, $D$ is not the most suitable distance in which the Hubble law is observationally preserved. More recently, however, it was demonstrated that the use of alternative versions of the luminosity distance are predictive and useful to determine viable priors which may calibrate the luminosity distance itself. A possible strategy consists in taking $D$ to limit $q_0,j_0,s_0,\ldots$ by fitting it with cosmic data. Afterwards, using $d_L$ in which $\Omega_k$ is not fixed as \emph{a priori}, it would be possible to infer limits on $\Omega_k$ itself, thereby fixing the cosmographic coefficients to the values determined through $D$. Another way out is to define it in terms of \emph{different} observable quantities which can be separately measured. For example, this may be determined by the expression:
\begin{equation}\label{equazioncontrdeg}
\Omega_k=\frac{\Big[H\,d_L^\prime\Big]^2-1}{\Big[H_0\,d_L\Big]^2}\,,
\end{equation}
with $d_L^\prime\equiv {d\,d_L\over dz} $. In this way, the spatial curvature is fixed by other sets of measurements and can be inserted in $d_L$, with the prescription that it does not vary significantly as the redshift increases in the small redshift regime.

\subsection{Dependence on the cosmological priors and the problem of truncated series}

Choosing the cosmological priors affects the numerical results of any cosmographic analysis, implying that the intervals of every cosmographic coefficient should be influenced correspondingly. The way to alleviate this issue may be naively imagined through enlarging the cosmological intervals over which one decides to span the cosmographic coefficients. However, a broadening of the systematics typically may occur, forcing one to employ hierarchical intervals in which different orders of cosmographic coefficients are investigated. To build up more stringent priors, one possibility is to calibrate the convergence ranges of each cosmographic parameter inside the $\Lambda$CDM Hubble sphere. If the coefficient ranges are outside the theoretical expectations, the corresponding prior would influence the analysis itself. In other words, if numerical outcomes provide good results with small errors, then one can suppose the priors are correctly defined. However, if this is not the case, then, depending on the error distribution, one can deduce two possibilities: $1)$ either that the $\Lambda$CDM model is not the correct late-time cosmological model\footnote{This is the case of small errors with cosmographic coefficients which depart from the theoretical priors.}, $2)$ or the experimental analyses is unable to constrain the cosmographic terms\footnote{This is the case of larger errors.}. Unfortunately, with current data the second case is the most likely one. To be more general, another assumption would be to take the $\Lambda$CDM model as a limiting case of a more general theory and so construct priors for $H_0,q_0$ from the $\Lambda$CDM model only, and then enlarge the priors for the remaining coefficients. In any case, constructing the cosmological priors in this way may influence the statistics. This happens because those $\Lambda$CDM-priors \emph{are centered} around the $\Lambda$CDM expectations. All the numerics would indicate it, rather than being fixed by it, otherwise no real departures from the $\Lambda$CDM model would be expected and the cosmographic approach fails to be predictive.

Typical intervals are:
\begin{subequations}\label{priors}
\begin{align}
  q_0 & \in[-0.95\,;\,-0.3]\,,\\
  j_0 & \in[0\,;\,2]\,,\\
  s_0 & \in[-2\,;\,7]\,,\\
  l_0 & \in[-5\,;\,10]\,,\\
  m_0 & \in[-10\,;\,50]\,,
\end{align}
\end{subequations}
In addition, evidence for slower convergence in the best fit algorithm leads to truncate series at a given order. Generally, enlarging the order corresponds to increasing systematics in the cosmographic measurements. However, lower orders badly influence the analysis because do not allow one to fix all the cosmographic coefficients. Again, in this case, one can adopt different orders of broadening samples, reducing the $N$-order parameter phase by fixing the $N-1$ coefficients through previous fits. This will also induce a correspondence between the increasing of the $N-1$ order coefficients with the $N$ parameters.

\subsection{Cosmography beyond general relativity}

The standard cosmological model assumes that a dark energy fluid speeds up the universe today. However another possibility is that the acceleration is driven by some type of modification of general relativity. It turns out that the construction of the cosmographic series depends upon on the choice of the particular gravitational theory. Consequently any deviations from general relativity would need different cosmographic definitions and the corresponding cosmographic series should be adjusted accordingly. This problem affects the reconstruction of dark energy models in both general relativity and extended theories of gravity. There exist issues if the cosmographic approach is applied to theories with higher derivatives in either the gravitational or matter sectors. Some authors claim that this is due to the fact that expansions of such models, such as $f(R)$, $f(T)$, Galileons, and so forth, lead to extra free parameters. Other authors assume a different perspective, by simply assuming that any cosmological model should be described by the $\Lambda$CDM model as the redshift approaches the present epoch. Let us consider two examples, concerning the class of scalar theories depending on a scalar field $\phi$ whose action reads: $\mathcal{S}=\int {\rm d}x^{4}\sqrt{-g}\left[ \frac{1}{2}R - \frac{1}{2} \omega (\phi)\partial_{\mu} \phi \partial^{\mu }\phi -V(\phi )+\mathcal{L}_{m}\right]$, with $\mathcal{L}_m$ the matter Lagrangian,  $\omega(\phi)$ the renormalizing factor and $V(\phi)$ the potential, at $z=0$, we have:
\begin{subequations}\label{ST7}
\begin{align}
\frac{V_0}{H_0^2}&=2-q_0-\frac{3\Omega_m}{2}\,,\\
\frac{V_{z0}}{H_0^2}&=4+3q_0-j_0-\frac{9\Omega_m}{2}\,,\\
\frac{V_{2z0}}{H_0^2}&=4+8q_0+j_0(4+q_0)+s_0-9\Omega_m\,,\\
\frac{V_{3z0}}{H_0^2}&=j_0^2-l_0-q_0j_0(7+3q_0)-s_0(7+3q_0)-9\Omega_m\,.
\end{align}
\end{subequations}
Without assuming that the universe today is modelled by $\Omega_m \approx 2/3(1+q_0)$, which represents a $\Lambda$ prior on the matter, the potential derivatives depend upon the mass and generally can degenerate with it. A more complicated example is offered by assuming the $f(R)$ case, with action $\mathcal{S}\,=\,\int \text{d}^4 x \sqrt{-g}\left[\frac{1}{2} f(R)+{\cal L}_{m}\right]\,.$, and at $z=0$:
\begin{subequations}
\begin{align}
\frac{f_0}{6H_0^2}&=-\alpha q_0 +\Omega_m + 6\beta\left( 2+q_0-j_0\right)\,,\\
\frac{f_{z0}}{6H_0^2}&=\,\alpha\left(2+q_0-j_0\right)\,,\\
\frac{f_{2z0}}{6H_0^2}&=\,6\beta\left(2+q_0-j_0\right)^2+\alpha\left[2+4q_0+(2+q_0)j_0+s_0\right]\,.
\end{align}
\end{subequations}
Here, the need of $\frac{{\rm d}f}{{\rm d}R}\big|_{R=R_0}=\alpha$ and $\frac{{\rm d}^2f}{{\rm d}R^2}\big|_{R=R_0}=\beta\,H_0^{-2}$ limits the analysis. A possible treatment is to assume the values of $\alpha=1$ and $\beta=0$ as a priori. This turns out to give pure general relativity at $z\rightarrow0$. If $\alpha$ and $\beta$ are taken free to vary, it is not reasonable to expect a one-to-one correspondence between the $f(R)$-derivatives and the cosmographic series. In all cases, the propagation of errors are often large and this frequently leads to degenerated fits. The use of cosmography therefore, has severe limitations in the field of modified gravity \cite{duns,orla0,orla01}. However, this does not mean that a cosmographic approach cannot be considered, together with other methods, as a way of ruling out different kinds of extended theories of gravity. More precisely, a fiducial model slightly different from $\Lambda$CDM may be derived if assumptions on the extra terms come from first principles, with proven mathematical validity. Hence, one would expect in such situations possible clear deviations away from the $\Lambda$CDM case, i.e., $(j_0=1)$. In other words, extra free parameters would not influence cosmography, as one fixes them from geometrical hypothesis or inside the theory itself and cosmography can be thought to hold also if the cosmographic series is not derived inside the modified theory itself.

\section{The convergence of cosmographic series: Auxiliary variables versus rational Pad\'e}

Up to now we have described the main problems common to all cosmographic approaches. All these issues are intimately related to each other and are most probably due to the fact that cosmography is essentially based on Taylor expansions, which are by definition approximations. In this section we discuss the most common cosmographic drawback, which deals with the fact that all our expansions lie on intervals which are close to $z\sim0$. Unfortunately, much of the available data lies outside this limit. Hence, theoretically speaking, it is unrealistic to use cosmic data in such intervals without expecting a bad convergence of cosmographic series. That is why all cosmographic analyses are deeply influenced by Taylor approximations and the quality of cosmographic reconstruction is seriously limited.

Mathematically speaking, the Taylor expansions converge if $z<1$, so that one would expect severe limitations on the convergence radii of any cosmographic analyses. One way to explore this issue is to assume auxiliary variables, hereafter $\mathcal{Z}_{new}$, which are constructed from functions of the redshift $z$, but provide  higher convergence radii\footnote{Auxiliary variables are also called \emph{alternative parameterisations}.}. The mathematical structure of such variables, here conventionally named $\mathcal{Z}_{new}$, might be derived from re-parameterising the redshift $z$, within a sphere of radius $\mathcal Z_{new}<1$ \cite{vysse}. Any auxiliary variable restricts such an interval in a shorter ensemble, having that:
\begin{subequations}\label{vvia}
\begin{align}
  z\rightarrow0 & \,\,\Rightarrow  \,\,\mathcal{Z}_{new}=0\,,\nonumber\\
  z\rightarrow\infty &\,\,\Rightarrow \,\,  \mathcal{Z}_{new}\leq1\,.
\end{align}
\end{subequations}
Furthermore, the construction of $\mathcal Z_{new}$ would require that:
\begin{itemize}
\item it must be feasible to invert it, passing from the redshift $z$ to it;
\item it might be one to one invertible with the redshift $z$, in the phase-space domain;
\item it should not diverge for any values of the redshift $z$ (for example it does not have to diverge at future times);
\item finally, any parametrisation needs to behave smoothly as the universe expands, without any critical points.
\end{itemize}
Ignoring one of the above conditions may lead to a misleading parametrisation which, although guaranteeing a redshift domain inside a larger convergence radius, does not lead to high-impact results. Furthermore, the choice of a new parametrisation becomes difficult to treat since it may be frequently plagued by biased estimators or by larger errors than using the standard $z$ redshift. A well-consolidate case is the first prototype of $\mathcal{Z}_{new}$, named the $y$-redshift and defined as:
\begin{equation}\label{y1}
y=\frac{z}{1+z}\,,
\end{equation}
As one can see, it provides a divergence at $z\rightarrow-1$ (future time), although it confines redshift data inside $y\in[0,1]$. Hence, from a theoretical point of view, our recipe suggests that $y$ is not the best choice in favour of $\mathcal{Z}_{new}$. Experiments seem to confirm this fact \cite{distanza,duns} and suggest that alternative versions of the auxiliary variables should be used. Common ones are given by\footnote{Here, we consider the notation $y_2$, $y_3$ and $y_4$ simply because we consider the $y$ redshift as $y=y_1$.}:
\begin{eqnarray}\label{ilbosco}
y_2&=&\arctan{\Bigl(\frac{z}{z+1}\Bigr)} = \arctan\left(1-a\right)\,,\nonumber\\
y_3&=&\frac{z}{1+z^2}=\frac{a}{(1-a)^{-1}-2a}\,,\\
y_4&=&\arctan{z}=\arctan{\left(\frac{1-a}{a}\right)}\,,\nonumber
\end{eqnarray}
whose limits are, for $z \in [0,\infty]: y_2\in [0,\frac{\pi}{4}], y_3\in [0,0], y_4\in [0,\frac{\pi}{2}]$ and $z \in [-1,0]: y_2\in
[\frac{\pi}{2},0], y_3\in [-\frac{1}{2},0], y_4\in [-\frac{\pi}{4},0]$ and in which we used the definition of the scale factor, i.e., $a\equiv(1+z)^{-1}$.
Even though appealing, the method of employing auxiliary variables suffers from the disadvantage of being an artificial trick, which does not introduce a new physical variable, but only a simple way to circumnavigate the convergence problem at high redshift. The differences between $z$ and auxiliary variables can be computed in various ways, not only through the basic demands that we invoked before. For example, focusing on $y$ only and building up {\em mock} data sets calibrated by the $\Lambda$CDM model with vanishing spatial curvature, one finds the results in the table below.
\begin{table}[htbp]
\caption{Tests for $\bm{\theta_1}$ and $\bm{\theta_2}$. We here report how many times the true parameters are inside the confidence levels. }
\label{tables1}
\begin{center}
\begin{tabular}{@{}cccccccccccccc@{}}
\hline
&     &     &     &  $\bm{\theta_1}$   &      &     &     &      &     &  $\bm{\theta_2}$   &      &     \\ \hline
& \vline    &   $y$   &   \vline  & \vline    &  $z$   &   \vline &   & \vline    &   $y$   &   \vline  & \vline    &   $z$   &   \vline  \\
\hline  & $1\sigma$ & $2\sigma$ &
 3$\sigma$ & $1\sigma$ & $2\sigma$ &
 3$\sigma$ &  & $1\sigma$ & $2\sigma$ &
 3$\sigma$ & $1\sigma$ & $2\sigma$ &
 3$\sigma$
\\ \hline
$q_0$ & 26 & 32 & 42 & 67   & 27 & 6 &  & 82 & 12 & 6 & 82   & 18 & 0  \\
$j_0$ & 10  &  45 & 45 & 64  & 29 & 7 &  & 93  &  5 & 2 & 88  & 12 & 0  \\
$s_0$ & 10  & 67 & 23 & 83  & 15 & 2 &  & 92  & 7 & 1 & 93  & 6 & 1 \\
$l_0$ & - & - & - & - & - & - &  & 100 & 0 & 0 & 100 & 0 & 0 \\
\hline
\end{tabular}
\end{center}
\end{table}
The table summarises the number of reasonable cosmographic results inside different confidence levels, by means of the union 2.1 data set, with a flat prior on magnitude errors of the form: $\sigma_{\mu}=0.15$. From the above results, by using both the redshift definitions, one can evaluate the cosmographic series and the corresponding posterior probabilities for each parameter\footnote{Here the analysis has been performed up to $l_0$, but results do not significantly change for further orders.}. The broadening order of hierarchy is given by $\bm{\theta_1} = \{H_0,q_0,j_0,s_0\}$ and $\bm{\theta_2} = \{H_0,q_0,j_0,s_0,l_0\}$, whereas $H_0$ is assumed to be marginalised. The constraints, performed for example using a Monte Carlo approach, are derived from the Mock data sets. They are distributed to suggest how frequent the true values are inside the $1,2,3-\sigma$ confidence levels. The redshift $z$ provides well-behaved coverage results, contrary to $y$ which manifests larger errors. This confirms the above theoretical description of auxiliary variables and puts severe limitations on their use.

A treatment, less artificial and motivated by the evident change of slope of magnitudes $\mu$ in the observable intervals of supernova data, is based on the concept of using rational approximations to describe the shape of universe's dynamics without considering truncated Taylor series in intervals exceeding the convergence radii. Among the various typologies of approximations \cite{barrow1}, an important role is played by the Pad\'e polynomials, which are defined as functions depending upon two parameters, $(n,m)$, given by the rational function \cite{padeoriginale}:
\begin{equation}\label{padedef}
P_{n m}(z)  = \frac{a_0+a_1\,z+\dots+a_n\,z^n}{1+b_1\,z+\dots+b_m\,z^m}\,,
\end{equation}
with degrees: $n\geq0$ (numerator) and $m\geq0$ (denominator). Assuming we wish to reconstruct the function $f(z)$, its derivatives at $z=0$ to the highest possible order should be built up by means of:
\begin{subequations}\label{padeeheh}
\begin{align}
P_{n m}(0)&=f(0)\,,\\
P_{n m}'(0)&=f'(0)\,,\\
P_{n m}^{m+n}(0)&=f^{m+n}(0)\,.
\end{align}
\end{subequations}
Pad\'e polynomials, sometimes improperly referred to as \emph{Pad\'e approximants}, for given $n$ and $m$ are unique up to an overall multiplicative constant.

Pad\'e approximations handle the divergence at high redshifts of the Taylor polynomials; in fact, while data taken over $z>1$ are quite unlikely to accurately fit Taylor series, the Pad\'e approach, instead, works fairly well, exactly for that regime \cite{gruberluongo}. Supposing one knows the magnitude shape estimate in two limiting cases, i.e., very small and very large redshifts respectively, we formally have:
\begin{subequations}\label{sus}
\begin{align}
d_L^0&\sim f_0+f'\,z+\frac{f''}{2}\,z^2\\
d_L^{\infty}&\sim g_0+g'\,\frac{1}{z}+\frac{g''}{2}\,\frac{1}{z^2}\,.
\end{align}
\end{subequations}
Thus, given a function that works as $d_L^0$ as $z \sim 0$, whereas $d_L^\infty$ as $z \sim \infty$, in both limits it converges to a finite value. It follows that the most natural function able to fit our numerics in both the two limits should be a \emph{rational function} of $z$. To this end, Pad\'e approximations represent a viable alternative to standard cosmography. For example, to quantify a possible advantage offered by them, one may evaluate the convergence radius $\mathcal R$ for different Pad\'e polynomials. For simplicity, assuming the case $n=1, m=1$, we have:
\begin{equation}
  \mathcal R_{\rm{Pad\acute{e}}} = \frac{2}{1-q_0} \,,
\end{equation}
testifying that for $q_0 > -1$
\begin{equation}
  \mathcal R_{\rm{Pad\acute{e}}}\geq \mathcal R_{\rm{Taylor}} \simeq 1\,.
\end{equation}
In addition, a qualitative improvement of adopting Pad\'e approximations is seen by working with the $\Lambda$CDM and $\omega$CDM models. For example, in the concordance paradigm, computing several $d_L$ Taylor and Pad\'e approximations for different orders provide the same results at small regimes. Moreover, it is useful to notice that third order Taylor polynomials are compared with $(1,1)$, $(1,2)$ and $(2,1)$ Pad\'e approximations while fourth degree Taylor expansions with $(1,3)$, $(3,1)$ and $(2,2)$ Pad\'e polynomials. Analogously for fifth order Taylor series we have $(1,4)$, $(3,2)$, $(2,3)$ and $(4,1)$ Pad\'e polynomials.

All those choices confirm the goodness of the Pad\'e approach but leaves unsolved the issue of degeneracy\footnote{Indeed, consider as an example the third order Taylor polynomials. They are approximated by $(1,2)$ and $(2,1)$ Pad\'e polynomials. So that, one has to justify which order is preferred than others.} between $(n,m)$. An empirical way is to match a theoretical model with Pad\'e polynomials and to check which ones are really favoured. For example, the Pad\'e polynomials $P_{22}$ and $P_{32}$  approximate the exact $\Lambda$CDM luminosity distance fairly well in the whole interval of redshift $z$. Moreover, it seems that in order to reproduce the late-time cosmology, the polynomials $P_{21}$, $P_{22}$ and $P_{32}$ best approximate the cosmic shape. This continues being valid in the case of the $\omega$CDM paradigm, which shows analogous results.

As a plausible theoretical scheme to choose the most viable Pad\'e orders, one may keep in mind \cite{luongopade,altropaddy}:
\begin{itemize}
  \item the Pad\'e function should smoothly evolve in all redshift ranges chosen for the specific cosmographic analysis\footnote{This might be guaranteed by the Pad\'e construction itself, with no singularities in the observable regime.};
  \item any Pad\'e approximation of the luminosity distance $d_{L}$ must be positive definite\footnote{otherwise the definition of magnitude would not hold at all.};
  \item the degree of the numerator and that of the denominator should be close, albeit with the prescription $n\lesssim m$;
  \item when using a particular data set or combined ones, the polynomials should be calibrated through viable cosmographic priors, which do not provide divergences.
\end{itemize}
The latter may induce bad convergence of Pad\'e approximations due to spurious divergences which arise in the Pad\'e $d_{L}$ due to the fact that they are fractional polynomials. In general, it is possible to overcome this issue simply building up appropriate Pad\'e polynomials.

\begin{figure}
\begin{center}
\includegraphics[width=3.4in]{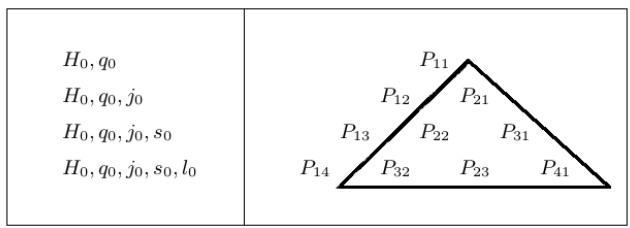}
{\small \caption{The Pad\'e approximates in the triangle give conclusive experimental results, while outside they seem to be un-predictive.}
\label{fig:PA}}
\end{center}
\end{figure}

We summarise some particular results on the Pad\'e best choices in Fig. \eqref{fig:PA}.

\section{The observational aspect of cosmography}

Throughout this review, we have investigated the role of cosmography from a theoretical point of view, analysing the main aspects of it, with particular attention given to the advantages and disadvantages of the cosmographic method. Even though our work represents a theoretical approach to cosmography, we should provide some thoughts on how to perform a numerical cosmographic analysis and how to handle some of the most common cosmic data sets. Our procedures should provide the best fit parameters with viable error estimations and then provide a way to evaluate how good the cosmographic fit is itself. Given the set of data points $\mathcal D_i$ and our cosmographic model, we define the $\chi$ function as follows \cite{licia}:
\begin{equation}\label{a1}
\chi^2=\sum_i w_i \mathcal F_i^2\,,
\end{equation}
where $w_i$ are suitably defined weights and $\mathcal F_i\equiv \mathcal D_i-y(x_i|\vec{\theta})$. If cosmic data are correlated then we formulate:
\begin{equation}\label{a2}
\chi^2=\sum_{ij} \Big[D_i-y(x_i|\theta)\Big]\,C_{ij}^{-1}\,\Big[D_j-y(x_j|\theta)\Big]\,,
\end{equation}
in which $C_{ij}$ is the covariance matrix. The best fit is the one which minimises the $\chi^2$. The quality of our data may be characterised by the quantity $C_{ij}^{-1}$. If the covariance matrix gives a very small probability, at the minimum of $\chi$, one may either have that the model does not work properly or the error distribution is non-Gaussian. A trick to check how well the cosmographic approach works is given by evaluating the pseudo-chi square, which is represented by the ratio of the standard chi square and the number of data points. This procedure can be adopted either for Gaussian or non-Gaussian distributions. The confidence limits are the regions in which the best fit parameters are confined, which may be represented by a compact region around the best set of coefficients. Assuming boundaries built up through  probability distributions, it is even possible to get ellipsoidal areas. Around the best fit those ellipses provide regions in which the chi square increases. Departures around the best chi square are imposed as $68.3\%$, $95.4\%$ and $99.5\%$ confidence levels, corresponding to $1$, $2$ and $3$-$\sigma$ respectively.

\subsection{The most common data sets}

In the context of cosmography the surveys of most interest are type Ia supernova observations. Over the years this data have been improved significantly and is used in many investigations. It is believed that these objects can be treated as standard candles, allowing one to get luminosity curves correlated with the distances of each supernova. One of the most important and recent supernova compilation is the union 2.1 catalogue \cite{dataunion}, which represents a well established extension of union and union 2 data sets~\cite{dataset1,dataset2}. Union 2.1 provides the great advantage to be systematics-free and so one would expect a net improvements of numerics through its use. Errors on the redshift are assumed to be negligibly small and so only the supernova magnitudes influence the cosmographic analysis itself. In this case, the chi square function is:
\begin{subequations}
\begin{align}
\chi^2_{SN} &= A - \frac{B^2}{C} + \log \left( \frac{C}{2\pi}\right)\,,\\
 A &= {\bf x}^T \mathcal{C}^{-1} {\bf x}\,, \nonumber \\
 B &= \sum_i (\mathcal{C}^{-1}{\bf x})_i\,,  \\
 C &= \text{Tr}[\,\mathcal{C}^{-1}\,]\,,  \nonumber
\end{align}
\end{subequations}
with $\mathcal{C}$ assumed here to be the covariance matrix and the $i$-th component of the vector ${\bf x}$:
\begin{equation}
 {\bf x}_i= 5\log_{10}\left(\frac{d_L(z_i; { \theta})}{\text{Mpc}}\right) + 25 - \mu_{obs}(z_i)\,.
\end{equation}

Another important data set stems from baryonic acoustic oscillations. Unfortunately, this measurement, albeit highly used in cosmology, is not perfectly model independent. In fact, it is weakly dependent on the $\Lambda$CDM model~\cite{d11}, since acoustic scales are functions of the drag time redshift, which may be obtained from a first order perturbation theory, once the minimal model is assumed. However, baryonic acoustic oscillations considered from~\cite{pxn2009} may be framed in terms of model independent quantity, $ \mathcal{B_R}\equiv D_V(0.35)/ D_V(0.20) = 1.736 \pm 0.065$, with volumetric distance given by $ D_V(z) = \left[c \,z\, d_L(z)^2/(1+z)^2H(z) \right]^{1/3}$. Here the $\chi$-squared function becomes
\begin{equation}
 \chi^2_{BAO}(\theta) = \frac{(D_V(0.35; \theta)/ D_V(0.20; \theta) - 1.736)^2}{0.065^2}\,.
\end{equation}
Another data set which makes use of 26 independent Hubble data points from~\cite{a,b,c,d,e,f,g}, and are the \emph{differential age measurements}. Here, massive early type galaxies are taken as cosmic chronometers~\cite{jy}. The strategy is to notice that one can evaluate $dt/dz$ which is related to the Hubble rate via the formula:
\begin{equation}\label{formula44}
H(z) = - (1+z)^{-1}dz/dt\,.
\end{equation}
The $\chi$-$squared$ function in this case is given by:
\begin{equation}
 \chi^2_{OHD}({\theta})  = \sum_{i} \frac{(H(z_i; {\theta}) - H_{obs}(z_i))^2 }{\sigma_i^2}\,.
\end{equation}
The last survey of data that we consider here are the amplitudes of the acoustic peaks in the cosmic microwave background (CMB) angular power spectrum. These mostly depend on the peak locations and on the matter density. The common use of the shift parameter, given by \cite{cmb}:
\begin{equation}\label{shift}
 R_{\mathrm{CMB}} = (\Omega_b + \Omega_{DM})^{0.5} \int_0^{z_{ls}} H_0 \frac{\mathrm{d} z'}{H(z')}\,,
\end{equation}
with $\Omega_b$ and $\Omega_{DM}$ respectively, baryonic and dark matter densities allows one to evaluate numerical outcomes making use of such a measurement. Here, $z_{ls}$ is the redshift at the last scattering surface. For unified cosmological models the sum of baryons and cold dark matter reduces to the baryon density only, i.e., $\Omega_b$, due to fact that the dark matter contributions naturally emerge as a consequence of dark energy. The fluid that permits the universe to accelerate at late times is able, in fact, to contribute to the total matter density, at early times. In other words, such a fluid behaves as a cosmological constant at present time and as a pressureless matter term at early times. In the case of unified models, we replace $R_{\mathrm{CMB}}$ through the ratio $2 l_1/l'_1$, where $l_1$ is a particular position of the first peak on the CMB, while $l'_1$ is the first peak in a Einstein-de Sitter universe.
In this case the chi square is:
\begin{equation}
 \chi^2 = \left( \frac{R_{CMB} - R_{CMB,obs}}{\sigma_R} \right)^2 \,.
\end{equation}
To conclude, due to the fact that the different data sets are uncorrelated, the total $\chi$ squares are given by the sum of all the involved chi squares.

\section{Concluding remarks and perspectives}

In this review, we have highlighted the main aspects of the cosmographic method, giving particular emphasis on its application to reconstructing the form of dark energy in a model independent way. In doing so, we proposed a set of basic strategies in order to build up the cosmographic method, introducing the concept of cosmological Taylor expansions and defining the cosmographic series. We then related the series of cosmographic coefficients to the derivatives of the Hubble rate and proposed how one might expand cosmological distances. Furthermore, we presented cosmographic definitions of the adiabatic sound speed, the energy  conditions, specific heats and the cosmological equation of state. These relations show how cosmography is able to reproduce constraints on the dark energy evolution, putting bounds on the thermodynamics of the fluid responsible for the cosmic speed up. At the level of the background cosmology, cosmography demonstrates that both the $\Lambda$CDM model and the dark fluid paradigm are favoured to describe the dynamics of the universe, simply because they reproduce more refined limits on the expansion history of the universe. The degeneracy between the two models is not completely removed through the cosmographic approach, although there are indications that an evolving dark energy contribution is currently favoured by present day cosmographic bounds.

We introduced the \emph{direct} and \emph{inverse} cosmographic procedures which aim to describe cosmography of generic cosmological models in two distinct cases. In the first, the direct approach, cosmographic bounds are inferred from rewriting the priors on the free parameters of a given model through the cosmographic series. In the second case, one obtains conditions on the free parameters, rewriting the luminosity distances in terms of them and then obtaining limits by means of cosmic data.

We then expressed three cosmological models of particular interest, i.e., the $\Lambda$CDM, $\omega$CDM and the CPL approach, in terms of the cosmographic series and showed how to match cosmography to them. All these properties and advantages of cosmography reproduce viable constraints but are unable to depict the whole expansion history of the universe, since several issues of using this approach plague the expected numerical outcomes. These problems have been interpreted in view of our considerations. In particular, we investigated the problem of degeneracy among cosmographic coefficients and how to alleviate it by using cosmological priors. We also focused on the degeneracy among coefficients and spatial curvature. In this case, we stressed that cosmography is essentially model independent only if $\Omega_k$ is somehow fixed or constrained. In doing so, we suggested strategies to either bound it from geometrical observations or from alternative techniques. Finally, we discussed the severe problem of handling cosmography with truncated series in general relativity or extended theories of gravity. In these cases, truncation at a given order may produce systematics leading to misleading results. Moreover, the dependence on the Friedmann equations limits the use of cosmography when investigating theories which depart from general relativity. We also demonstrated that the use of auxiliary variables are a viable possibility in order to elevate these problems, although the best way of building up the \emph{most suitable} $z$-parametrisation is not precisely known. Consequently, in order to go beyond this problem we described the use of Pad\'e approximations in the field of cosmology. We showed that its use may overcome the problems due to truncated series, since the Pad\'e polynomials match data at higher redshift intervals, without the need of rewriting the redshift variables using alternative parameterisations. In doing so, we faced the more general problem of convergence which affects cosmography and does not permit one to use it as the definitive approach to describe the expansion history of the universe with high precision. We also discussed a number of experimental problems, describing briefly how one uses the Bayesian approach and Monte Carlo techniques to build up cosmographic observations. Finally we defined the most common chi squared definitions, describing the common data sets used in the literature. We believe that future work needs to focus on applications of cosmography to higher redshift domains. This can be done through the use of rational reconstructions of the cosmographic series and/or by means of further data. Moreover, possible future applications of cosmography in the limit of perturbation theory are essential to connect both late and early times of the universe. Summing up, the role of cosmography may open new insights on the evolution of the universe. The present state of the art focuses on the late (low-redshift) universe only, but upcoming work will focus on obtaining a high-redshift cosmography with increasing accuracy.


\appendix

\section{Appendix A: Cosmographic series of the principal cosmological models}

\subsection{The $\Lambda$CDM case}

The cosmographic series for the concordance model, up to the snap parameter, is given by:

\begin{subequations}\label{lfcosmo1}
\begin{align}
q_0&= \frac{3}{2}\Omega_m -1+\Omega_k\,,\\
j_0&= 1-\Omega_k\,,\\
s_0&=(1-\Omega_k)^2-\frac{3}{2}\Omega_m\left(3-5\Omega_k\right)\,.
\end{align}
\end{subequations}

\subsection{The $\omega$CDM case}

The cosmographic series for the $\omega$CDM model, up to the snap parameter, is given by:

\begin{subequations}\label{lfcosmo2}
\begin{align}
q_0&= \frac{1}{2} \Big[1 - \Omega_k - 3 (-1 + \Omega_k + \Omega_m)\omega\Big]\,,\\
j_0&= \frac{1}{2}\Big\{2\Omega_m-(-1+\Omega_k+\Omega_m)\big[2+9\omega(1 + \omega)\big]\Big\}\,,\\
s_0&= \frac{1}{4}\Big\{2(7 - \Omega_k) (-1 + \Omega_k) - 3 (-27 + 5\Omega_k) (-1 + \Omega_k + \Omega_m) \omega - 9 (-1 + \Omega_k + \Omega_m) (-16 + 4\Omega_k + 3\Omega_m) \omega^2\nonumber\\ &-27 (-3 + \Omega_k + \Omega_m) (-1 + \Omega_k + \Omega_m) \omega^3\Big\}\,.
\end{align}
\end{subequations}

\subsection{The CPL case}

The cosmographic series for the CPL model, up to the snap parameter, is given by:

\begin{subequations}\label{lfcosmo3}
\begin{align}
q_0&=\frac{1}{2}[1 - \Omega_k - 3 (-1 + \Omega_k + \Omega_m) \omega_0]\,,\\
j_0&=\frac{1}{2}\Big\{2\Omega_m - (-1 + \Omega_k + \Omega_m) [2 + 9 \omega_0(1 + \omega_0) + 3 \omega_1]\Big\}\,,\\
s_0&=\frac{1}{4} \{-14-\Omega_k(1+3\omega_0)(-16+3\omega_0(-16+5\Omega_m+6(-2+\Omega_m)\omega_0))-\nonumber\\&9(-1+\Omega_m)\omega_0(-9+\omega_0(-16-9\omega_0+       3\Omega_m(1+\omega_0)))-\nonumber\\&33\omega_1-3\Omega_k(-12+\Omega_m+6(-4+\Omega_m)\omega_0)\omega_1+\nonumber\\&3(11\Omega_m-3(-7+\Omega_m)(-1+\Omega_m)\omega_0)\omega_1 -\Omega_k^2(1 + 3\omega_0)(2+9\omega_0(1 + \omega_0)+3\omega_1)\}\,.
\end{align}
\end{subequations}


\begin{thebibliography}{99}

\bibitem{riess}
A. G. Riess, et al., Astron. J., {\bf 116}, 1009-1038, (1998); J. L. Tonry, et al., Astrophys. J., {\bf 594}, 1-24, (2003); A. G. Riess, et al., Astrophys. J., {\bf 607}, 665-687, (2004).

\bibitem{perlmutter}
S. Perlmutter, et al., Astrophys. J., {\bf 517}, 565-586, (1999); N. A. Bahcall, J. P. Ostriker, S. Perlmutter, P. J. Steinhardt, Science, {\bf 284}, 1481-1488, (1999); R. A. Knop, et al., Astrophys. J., {\bf 598}, 102, (2003).

\bibitem{miao}
M. Li, X.D. Li, S. Wang, Y. Wang, Front. Phys. {\bf 8}, 828-846, (2013); V. Sahni, Lect. Notes Phys., {\bf 653}, 141-179, (2004).

\bibitem{copeland}
J. E. Copeland, M. Sami, S. Tsujikawa, Int. J. Mod. Phys. D, {\bf 15}, 1753-1936, (2006).

\bibitem{extended}
S. Capozziello, M. De Laurentis, Phys. Rept., {\bf 509}, 167-321, (2011); S. Capozziello, M. De Laurentis, V. Faraoni, Open Astron. J., {\bf 3}, 49, (2010).

\bibitem{ioequevedo}
O. Luongo, H. quevedo, Phys. Rev. D, {\bf 90}, 8, 084032, (2014); O. Luongo, H. Quevedo, ArXiv[gr-qc]:1507.06446, (2015); O. Luongo, H. Quevedo, Proc. MG XII, Paris, DOI:$10.1142/9789814374552_0122$, (2011); J. W. Maluf, Gen. Rel. Grav., {\bf 46}, 1734, (2014).

\bibitem{darkenergy}
T. Padmanabhan, AIP Conf. Proc., {\bf 861}, 179-196, (2006); J. L. Cervantes-Cota, G. Smoot, AIP Conf. Proc., {\bf 1396}, 28-52, (2011); E. V. Linder, R. J. Scherrer, Phys. Rev. D, {\bf 80}, 023008, (2009).

\bibitem{ilmodello}
S. Weinberg, Rev. Mod. Phys., {\bf 61}, 1, (1989); T. Padmanabhan, Phys. Rept., {\bf 380}, 235-320, (2003).

\bibitem{ilmiopaperconquew}
O. Luongo, H. Quevedo, Int. J. Mod. Phys. D, {\bf 23}, 1450012, (2014).

\bibitem{darkfluid}
M. Kunz, Phys. Rev. D, {\bf 80}, 123001, (2009).

\bibitem{losdos}
P. J. E. Peebles, B. Ratra, Rev. Mod. Phys., {\bf 75}, 559-606, (2003).

\bibitem{iniziocosmografia}
S. Weinberg, \emph{Gravitation and Cosmology: Principles and Applications of the General Theory of Relativity}, Wiley, Hoboken, NJ, USA, (1972); M. Visser, Phys. Rev. D, {\bf 56}, 7578, (1997); M. Visser, Science, {\bf 276}, 88, (1997).

\bibitem{harrison}
E. R. Harrison, Nature, {\bf 260}, 591, (1976);

\bibitem{noein}
M. Visser, Gen. Rel. Grav., {\bf 37}, 1541-1548, (2005).

\bibitem{noein2}
K. Lake, Phys. Rev. D, {\bf 76}, 063508, (2007).

\bibitem{noein3}
C. Cattoen, M. Visser, Class. Quant. Grav., {\bf 25}, 165013, (2008); C. Cattoen, Class. Quant. Grav., {\bf 24}, 5985-5998, (2007).

\bibitem{fluidi}
J. L. Cervantes-Cota, J. Klapp, Springer series title 7487, 71-105, (2014); K. Bamba, S. Capozziello, S. Nojiri, S. D. Odintsov, Astrophys. Space Sci., {\bf 342}, 155-228, (2012).

\bibitem{vi2005}
M. Visser, Class. Quant. Grav., {\bf 32}, 13, 135007, (2015).

\bibitem{ade}
P. A. R. Ade, {\it et al.}, [Planck Collaboration], Astron. Astrophys., A, {\bf 16}, 571, (2014).

\bibitem{noosi}
A. C. C. Guimares, J. V. Cunha, J. A. S. Lima, JCAP, {\bf 0910}, 010, (2009);  M. Visser, C. Cattoen, Proceedings of DARK 2009, The Seventh Intern. Heidelberg Conf. on Dark Matter,  Astro and Particle Physics, Christchurch, New Zealand, (2009); R. Cai, Z. Tuo, Phys. Lett. B, {\bf 706}, 116-122, (2011).

\bibitem{nohq}
A. R. Neben, M. S. Turner, ApJ, {\bf 769}, 133, (2013).


\bibitem{ohd}
M. Scrimgeour et al., Mon. Not. Roy. Astron. Soc., {\bf 425}, 116, (2012); E. Caztanaga, A. Cabre and L. Hui, Mon. Not. Roy. Astron. Soc., {\bf 399}, 1663, (2009).

\bibitem{hubblerate1}
O. Luongo, Mod. Phys. Lett. A, {\bf 26}, 1459-1466, (2011).

\bibitem{hubblerate2}
L. Xu, Y. Wang, Phys. Lett. B, {\bf 702}, 114-120, (2011).

\bibitem{distanza}
A. Aviles, C. Gruber, O. Luongo, H. Quevedo, Phys. Rev. D, {\bf 86}, 123516, (2012).


\bibitem{iosuono2}
O. Luongo, H. Quevedo,  Astroph. sp. sci., {\bf 338}, 2, 345-349, (2012).

\bibitem{equax}
S. Capozziello, M. De Laurentis, O. Luongo, A. C. Ruggeri, Galaxies, {\bf 1}, 216-260, (2013).

\bibitem{cattviss2007}
C. Cattoen, M. Visser, ArXiv[gr-qc]:0703122, (2007).

\bibitem{cattviss2005}
C. Cattoen, M. Visser, Class. Quant. Grav., {\bf 22}, 4913-4930, (2005).

\bibitem{weinberg2008}
S. Weinberg, \emph{Cosmology}, (Oxford Univ. Press, Oxford), (2008).

\bibitem{annalen}
S. Capozziello, M. De Laurentis, O. Luongo, Annalen Phys., {\bf 526}, 309-317, (2014).

\bibitem{hung8}
R. Nair, S. Jhingan, and D. Jain, JCAP {\bf 1105}, 023, (2011); B. A. Bassett and M. Kunz, Phys. Rev. D {\bf 69}, 101305, (2004).

\bibitem{bll}
I. M. H. Etherington, Gen. Rel. Grav. {\bf 39}, 1055, (2007); R. Kristian and R. K. Sachs, Astrophys. J., {\bf 143}, 379 (1966).


\bibitem{bll3}
R. F. L. Holanda, J.A.S. Lima, and M. B. Ribeiro, Astron. Astrophys. {\bf 528}, L14 (2011);
F. De Bernardis, E. Giusarma, and A. Melchiorri, Int. J. Mod. Phys. D {\bf 15}, 759 (2006).

\bibitem{bll4}
G. F. R. Ellis, Gen. Rel. Grav. {\bf 39}, 1047, (2007).

\bibitem{carroll}
S. Carroll, \emph{Spacetime and geometry}, (Addison-Wesley), Pearson, (2004).

\bibitem{defsound}
S. DeDeo, R. R. Caldwell, and P. J. Steinhardt, Phys. Rev. D, {\bf 67}, 103509, (2003); R. Bean and O. Dore, Phys. Rev. D, {\bf 69}, 083503, (2004); J. Weller and A. M. Lewis, Mon. Not. Roy. Astron. Soc., {\bf 346}, 987–993, (2003); S. Hannestad, Phys. Rev. D, {\bf 71}, 103519, (2005); P. S. Corasaniti, T. Giannantonio, and A. Melchiorri, Phys. Rev. D, {\bf 71}, 123521, (2005); J. Q. Xia, Y. F. Cai, T. T. Qiu, G. B. Zhao, and X. Zhang, Int. J. Mod. Phys. D, {\bf 17}, 1229-1243, (2008); D. F. Mota, J. R. Kristiansen, T. Koivisto, and N. E. Groeneboom, 	Mon. Not. Roy. Astron. Soc., {\bf 382}, 793-800, (2007); R. de Putter, D. Huterer, and E. V. Linder, Phys. Rev. D, {\bf 81}, 103513, (2010); H. Li and J. Q. Xia, JCAP, {\bf 1004}, 026, (2010).

\bibitem{ellyx}
G. Ellis, R. Maartens, M. MacCallum, Gen. Rel. Grav., {\bf 39}, 1651, (2007).

\bibitem{darkma}
J. Barranco, A. Bernal, and D. Nunez, Mon. Not. Roy. Astron. Soc., {\bf 449}, 1, 403-413, (2015).

\bibitem{suono3}
L. Xu, Phys. Rev. D., {\bf 87}, 043503, (2013); L. Xu, Y. Wang, H. Noh, Phys. Rev. D, {\bf 85}, 043003, (2012); Alexandre Arbey, AIP Conf. Proc.,  {\bf 1241}, 700-707, (2010).

\bibitem{Hu:1998tj}
W. Hu and D. J. Eisenstein, Phys. Rev. D, {\bf 59}, 083509, (1999).

\bibitem{debate}
M. Nouri-Zonoz, J. Koohbor, H. Ramezani-Aval, Phys. Rev. D, {\bf 91}, 063010, (2015); W. Yang, L. Xu, Y. Wang, Y. Wu, Phys. Rev. D, {\bf 89}, 043511, (2014); M. Kunz, A. R Liddle, D. Parkinson, C. Gao, Phys. Rev. D, {\bf 80}, 083533, (2009); J. L. Cervantes-Cota, A. Aviles, J. De-Santiago, AIP Conf. Proc., {\bf 1548}, 299-313, (2013).

\bibitem{nosoundzero}
N. Bilic, G. B. Tupper, R. D. Viollier, Phys. Lett. B, {\bf 535}, 17-21, (2002); A. Aviles, A. Bastarrachea-Almodovar, L. Campuzano, H. Quevedo, Phys. Rev. D, {\bf 86}, 063508, (2012);

\bibitem{nojiri}
S. Nojiri, S. D. Odintsov, Phys. Lett. B, {\bf 649}, 440-444, (2007); A. Arbey, Phys. Rev. D, {\bf 74}, 043516, (2006).

\bibitem{Aviles:2011ak}
A. Aviles, J. L. Cervantes-Cota, Phys. Rev. D, {\bf 84}, 083515, (2011).

\bibitem{pad}
T. Padmanabhan, Rept. Prog. Phys., {\bf 73}, 046901, (2010); D. Kothawala, S. Sarkar, T. Padmanabhan, Phys. Lett. B, {\bf 652}, 338-342, (2007).

\bibitem{herny}
A. Krasinski, H. Quevedo, R. Sussman, J. Math. Phys., {\bf 38}, 2602, (1997).

\bibitem{ioavileseklapp}
A. Aviles, N. Cruz, J. Klapp, O. Luongo, Gen. Rel. Grav., {\bf 47}, 5, 63, (2015).

\bibitem{rubano}
C. Rubano, P. Scudellaro, Gen. Rel. Grav., {\bf 34}, 1931-1939, (2002).

\bibitem{lib12}
N. R. Draper, H. Smith, \emph{Applied Regression Analysis}, (Wiley, New York), (1998); P. Bevington, et al., \emph{Data Reduction and Error Analysis for the
Physical Sciences}, (McGraw-Hill), (2002).

\bibitem{deg88}
I. Wasserman, Phys. Rev. D, {\bf 66}, 123511, (2002); J. L. Crooks, J. O. Dunn, P. H. Frampton, H. R. Norton, T. Takahashi, Astropart. Phys., {\bf 20}, 361-367, (2003).

\bibitem{gruberluongo}
C. Gruber, O. Luongo, Phys. Rev. D, {\bf 89}, 103506, (2014).

\bibitem{ricostruzione1}
T. Holsclaw, U. Alam, B. Sanso, H. Lee, K. Heitmann, S. Habib, D. Higdon, Phys. Rev. Lett., {\bf 105}, 241302, (2010); T. Holsclaw, U. Alam, B. Sanso, H. Lee, K. Heitmann, S. Habib, D. Higdon, Phys. Rev. D, {\bf 84}, 083501, (2011); P. P. Avelino, L. Losano, R. Menezes, J. C. R. E. Oliveira, Phys. Lett. B, {\bf 717}, 313-318, (2012).

\bibitem{ricostruzione2}
A. Shafieloo, V. Sahni, A. A. Starobinsky, Phys. Rev. D, {\bf 86}, 103527, (2012); C. Clarkson, M. Cortes, M. and B. J. Bassett, Cosm. Astrop. Phys., {\bf 8}, 11,  (2007); J. A. Vazquez, M. Bridges, M. P. Hobson, A. N. Lasenby, JCAP, {\bf 020}, 1209, (2012); M. Seikel, C. Clarkson, M. Smith, JCAP, {\bf 06}, 036, (2012).

\bibitem{yes}
E. Bianchi, C. Rovelli, {\emph Why all these prejudices against a constant?}, ArXiv[astro-ph.CO]:1002.3966, (2010); C. J. A. P. Martins, AIP Conf. Proc., {\bf 105}, 1514, (2012).

\bibitem{ornot}
V. Pavlidou, T. N. Tomaras, JCAP, {\bf 020}, 1409, (2014); M. Kunz, J. Phys. Conf. Ser., {\bf 110}, 062014, (2007).

\bibitem{cp}
M. Chevallier, D. Polarski, Int. J. Mod. Phys. D., {\bf 10}, 213–224, (2001).

\bibitem{elle}
E. V. Linder, Phys. Rev. Lett., {\bf 90}, 091301, (2003).

\bibitem{chris2007}
C. Clarkson, M. Cortes, B. A. Bassett, JCAP, {\bf 011}, 0708, (2007); A. Pavlov, S. Westmoreland, K. Saaidi, B. Ratra, Phys. Rev. D, {\bf 88}, 123513 (2013).

\bibitem{duns}
V. C. Busti, P. K. S. Dunsby, A. de la Cruz-Dombriz, D. Saez-Gomez, \emph{Is cosmography a useful tool for testing cosmology?}, ArXiv[astro-ph.CO]:1505.05503, (2015).

\bibitem{orla0}
A. Aviles, A. Bravetti, S. Capozziello, O. Luongo, Phys. Rev. D, {\bf 87}, 044012, (2013).

\bibitem{orla01}
A. Aviles, A. Bravetti, S. Capozziello, O. Luongo, Phys. Rev. D, {\bf 87}, 064025, (2013).

\bibitem{vysse}
M. Visser, Class. Quant. Grav., {\bf 21}, 2603-2616, (2004).

\bibitem{barrow1}
G. A. Baker, jr. and P. Graves, \emph{Encyclopedia of Mathematics and its Applications}, volumes 13 and 14, (Addison-Wesley), (1981); W. H. Press, S. A. Teukolsky, W. T. Vetterling and B. P. Flannery, \emph{Numerical Recipes}, (Cambridge University Press, Cambridge), (1993); A. R. Liddle, P. Parsons, J. D. Barrow, Phys. Rev. D, {\bf 50}, 7222-7232, (1994).

\bibitem{padeoriginale}
M. Della Morte, B. Jager, A. Juttner and H. Wittig, JHEP, {\bf 055}, 1203, (2012); W. B. Jones and W. J. Thron, \emph{Continued fractions,  Analytic theory and applications}, volume 11 of Encyclopedia of Mathematics and its Applications, (Addison-Wesley Publishing Co.), (1980); S. G. Krantz and H. R. Parks, \emph{A primer of real analytic functions}, (Birkauser), (1992).

\bibitem{luongopade}
A. Aviles, A. Bravetti, S. Capozziello, O. Luongo, Phys. Rev. D, {\bf 90}, 043531, (2014).

\bibitem{altropaddy}
H. Wei, X.-P. Yan, Y.-N. Zhou, JCAP, {\bf 045}, 1401, (2014); J. Liu, H. Wei, \emph{Cosmological Models and Gamma-Ray Bursts Calibrated by Using Pad\'e Method}, ArXiv[astro-ph.CO]:1410.3960, (2014); A. Alho, C. Uggla, J. Math. Phys., {\bf 56}, 012502, (2015); I. Semiz, A. K. Camlibel, \emph{What do the cosmological supernova data really tell us?}, ArXiv [gr-qc]:1505.04043, (2015).

\bibitem{licia}
L. Verde, Lect. Not. Phys. {\bf 800}, 147-177, (2010).

\bibitem{dataunion}
N. Suzuki, et al., Astrophys. J., {\bf 85}, 746, (2012).

\bibitem{dataset1}
M. Kowalski, et al., Astrophys. J., {\bf 686}, 749, (2008).

\bibitem{dataset2}
R. Amanullah, et al., Astrophys. J., {\bf 716}, 712, (2010).

\bibitem{d11}
R. Durrer, Philos. Trans. R. Soc. A, {\bf 369}, 5102, (2011).

\bibitem{pxn2009}
W. J. Percival, et al., Mon. Not. R. Astron. Soc., {\bf 401}, 2148, (2010).

\bibitem{a}
J. Simon, L. Verde, R. Jimenez, Phys. Rev. D, {\bf 71}, 123001, (2005).

\bibitem{b}
D. Stern, et al., JCAP, {\bf 008}, 1002, (2010).

\bibitem{c}
M. Moresco, et al., JCAP, {\bf 006}, 1208, (2012).

\bibitem{d}
N.G. Busca, et al. Astr. $\&$ Astroph., {\bf 552}, A96, (2012).

\bibitem{e}
C. Zhang, et al., Res. Astron. and Astroph., {\bf 14}, 1221, (2014).

\bibitem{f}
C. Blake, et al., MNRAS, {\bf 425}, 405, (2012).

\bibitem{g}
C.H. Chuang, et al., MNRAS, {\bf 435}, 1, 255, (2013).

\bibitem{jy}
R. Jimenez, et al. ApJ, {\bf 593}, 622, (2003).

\bibitem{cmb}
L. Verde, H. Peiris, R. Jimenez, JCAP, {\bf 019}, 0601, (2006).

\end{thebibliography}
\end{document}